\documentclass[useAMS,usenatbib]{mnras}
\usepackage[fleqn]{amsmath}
\usepackage{graphicx}	
\pdfoutput=1 

\newcommand{\der}{{\rm d}}

 \newcommand{\rel}{_{\rm rel}}
 \newcommand{\acc}{^{\rm acc}}

 \newcommand{\h}{_{\rm h}}

 \newcommand{\clR}{R_{\rm s}} 
\newcommand{\clM}{M_{\rm s}}

 \newcommand{\tr}{^{\rm tr}} \newcommand{\trb}{_{\rm rel} }

 \newcommand{\dc}{_{\rm dDM}}

\newcommand{\rs}{r_{\rm s}} 
 
\newcommand{\modot}{M$_\odot$\ } \newcommand{\modotc}{M$_\odot$}

\newcommand{\beq}{\begin{equation}} \newcommand{\eeq}{\end{equation}}
\newcommand{\med}{_{\rm med}}
 \newcommand{\beqa}{\begin{eqnarray}}
\newcommand{\eeqa}{\end{eqnarray}} \newcommand{\lav}{\langle}
\newcommand{\rav}{\rangle}

 \newcommand{\f}{_{\rm f}} \newcommand{\PA}{^{\rm PA}}  
 
 \newcommand{\uc}{^{\rm c}}
\newcommand{\fin}{^{\rm stp}}

\begin{document}

\title[The Mass and Formation Time of the Host Haloes]{An Accurate Comprehensive Approach to Substructure: \\III. Masses and Formation Times of the Host Haloes}

\author[Salvador-Sol\'e, Manrique, Canales \& Botella]{Eduard
  Salvador-Sol\'e$^1$\thanks{E-mail: e.salvador@ub.edu}, Alberto Manrique$^1$, David Canales$^2$ and Ignacio Botella$^{1,3}$
  \\$^1$Institut de Ci\`encies del Cosmos, Universitat de
  Barcelona, Mart{\'\i} i Franqu\`es 1, E-08028 Barcelona, Spain
  \\$^2$School of Aeronautics and Astronautics, Purdue University, 701 W. Stadium Ave., West Lafayette, IN 47907-2045, USA
  \\$^3$Dept. of Astronomy, Graduate School of Science, Kyoto University, Kitashirakawa, Oiwakecho, Sakyo-ku, Kyoto, 606-8502, Japan}


\maketitle
\begin{abstract}
With this Paper we complete a comprehensive study of substructure in dark matter haloes. In Paper I we derived the radial distribution and mass function (MF) of accreted subhaloes (scaled to the radius and mass of the host halo) and showed they are essentially universal. This is not the case, however, for those of stripped subhaloes, which depend on halo mass and assembly history. In Paper II we derived these latter properties in the simplest case of purely accreting haloes. Here we extend the study to ordinary haloes having suffered major mergers. After showing that all the properties of substructure are encoded in the mean truncated-to-original subhalo mass ratio profile, we demonstrate that the dependence of the subhalo MF on halo mass arises from their mass-dependent concentration, while the shape of the subhalo radial distribution depends on the time of the last major merger of the host halo. In this sense, the latter property is a better probe of halo formation time than the former. Unfortunately, this is not the case for the radial distribution of satellites as this profile is essentially disconnected from subhalo stripping and the properties of accreted subhaloes are independent of the halo formation time. 
\end{abstract}



\section{INTRODUCTION}\label{intro}

In the last decade, the subject of halo assembly history has attracted much attention in connection with the so-called ``missing satellite problem", namely that the abundance of satellite galaxies in the Milky Way (MW) (and Andromeda; \citealt{Tea14}) does not seem to conform with the expected one in the favourite $\Lambda$CDM cosmology \citep{Mea99,Kea99,BB17}. Indeed, one possible explanation for that problem is that the MW halo may not have had the typical assembly history of haloes of its mass. Some aspects of the MW suggest, indeed, that it has had a particularly quiescent history (e.g. \citealt{Wy01,De13,Ru15,La19}). 

Substructure in dark matter haloes is believed to harbour important information on their assembly history as it is the direct consequence of the way they have grown. Indeed, haloes undergo long periods of frequent {\it minor} mergers (generically called accretion), separated by sporadic {\it major} mergers. The difference between these two kinds of mergers is that the largest halo in minor mergers (the accreting object) is much more massive than its partners (the accreted objects), so it remains essentially in equilibrium during that process and the accreted haloes survive within their host as subhaloes. On the contrary, all haloes (usually two) partaking of a major merger are similarly massive so that the event causes them to go out of equilibrium and to form a new virialised halo after the system relaxes again. In this case the progenitor haloes are thus destroyed, but their subhaloes are transferred to the newborn halo. Subhaloes thus accumulate within haloes, where they are more or less stripped and shock-heated depending on the characteristics of each transient host. Consequently, their final properties are the result of their past history.

Unfortunately, we do not know yet what are the typical properties of substructure in haloes of different masses and formation times. None of the analytic models of halo substructure so far developed (\citealt{TB01,ZB03,S03,L04,OL04,TB04,Pe05,vdB05,Zeabis05,KB07,Gi08,Bea13,PB14,Ji16,Gfea16,vdB16,vdB18,GB19,Fea20}) have been able to provide definite answers to these questions. Nor have high-resolution simulations, which have only been able to draw the properties of substructure in a handful of haloes of the MW-mass \citep{Dea07,Sea08,BK10,Wa11} or of other masses (e.g. \citealt{Aea09,Eea09,Gi10,Kea11,Gea11,Gea12,Oea12,Lea14,Cea14,Iea20,Lea21}). 

But things are rapidly changing. The incoming new data gathered by means of the {\it Gaia} satellite \citep{Ga18} will allow to accurately determine the MW's substructure and it has now become feasible to observe satellites in neighbouring MW analogues \citep{Da17,Ge17,Sm18,Be19,Be20,Cr19,Ca20,Ma20,Ca21}. In parallel, simulations have also greatly improved. The {\it Copernicus Complexio} $N$-body simulations together with a semi-analytic galaxy formation model (\citealt{Hell16,Bea16,Bea20}) or the {\it Apostle} and {\it Auriga} \citep{Rich20}, {\it FIRE-2} \citep{Sam20}, and {\it Artemis} \citep{Fea20,Fea21,Eea21} hydrodynamic simulations have gathered a considerable number of simulated MW (and Andromeda) analogs. On the other hand, it is now possible to reach very high stellar mass resolutions \citep{Gea21} which significantly improves the statistics of substructure at the level of faint and ultra-faint satellites ($10^2$ \modot $<M_\star<10^6$ \modotc).

In addition, great progress has also been made on analytic grounds. A very complete model of substructure formation has been built \citep{Jea21} that allows one to study the effects of different initial conditions in the accretion and evolution of satellite galaxies. Likewise, the powerful {\it ConflUent System of Peak trajectories} (CUSP) formalism \citep{Mea95,Mea96,Mea98,Sea12a,Sea12b,Jea14a,Jea14b}, making the link between the properties of haloes and their seeds (peaks) in the random Gaussian field of density perturbations (see \citealt{SM19} for an overview) has been successfully applied to the study of halo substructure (\citealt{I} and \citealt{II}, hereafter Papers I and II, respectively).

In Paper I we determined the properties of accreted subhaloes, which act as initial conditions in their evolution through stripping inside their host haloes. To do this we took profit of the fact that, as shown in \citet{Jea19}, all halo properties arising from their gravitational clustering process do not depend on their particular assembly history so that one has the right to focus on the simplest case of {\it purely accreting haloes} evolving inside-out (see below for a brief explanation of this important result). However, subhalo stripping as well as dynamical friction are two (coupled) mechanisms acting on the dynamical evolution of subhaloes that are not directly connected to gravitational clustering. Consequently, the properties of stripped subhaloes do depend on the halo assembly history. In Paper II we built a detailed model of tidal stripping and shock heating of subhaloes as they orbit inside haloes which allowed us to derive their final properties. However, this was done in the simplest case of purely accreting haloes only and neglecting dynamical friction. 

In the present Paper, we complete this study and extend the characterisation of substructure to ordinary haloes, i.e. haloes having suffered major mergers, paying special attention to the role of halo mass and formation time and analysing the possible use of those properties as a probe for halo assembly history. Our treatment does not include dynamical friction. However, the predictions for low-mass subhaloes ($\clM\la 10^{-4}M\h$, where $M\h$ is the mass of the host halo) should not be affected by that omission. On the other hand, it does not include baryons either. However, by comparing our predictions to the results of simulations including them, it is still possible to unravel to some extent the influence of baryons physics in the properties of substructure. 

The layout of the Paper is as follows. In Section \ref{PA} we remind the main results of Papers I and II for purely accreting haloes of different masses. In Section \ref{ordinary} we extend those results to haloes having suffered major mergers. And in Section \ref{specific} we analyse the properties of substructure in haloes of a fixed mass and different formation times. Our results are summarised and discussed in Section \ref{dis}.

Some comments on the notation used in this Paper are in order. Unless otherwise stated, when we refer to subhaloes without specifying their kind, we mean stripped (or truncated) subhaloes. The halo formation time used is defined as the time they suffered their last major merger. It thus differs from the most usual definition: the time haloes reach 50\% of their final mass. We prefer the former not only because it is less arbitrary (why 50\% and not, say, 75\%?), but also because it is physically better motivated. Indeed, as mentioned, virialised haloes interrupt their identity in major mergers where they are destroyed and a new virialised object appears \citep{SM19}. Lastly, the notation we use for the cumulative or differential abundances of subhaloes, dependent in general on subhalo mass, $\clM$, and radial location inside the host halo, $r$, is the same as in Papers I and II. The cumulative number of stripped or accreted haloes out to $r$ and down to $\clM$ are denoted as ${\cal N}\fin(< r, >\clM)$ and ${\cal N}\acc(< r, >\clM)$, respectively. When one of the arguments takes its maximum value, i.e. when the integrals over $r$ or $\clM$ are complete, we drop the corresponding argument. For instance, ${\cal N}\fin(>\clM)$ stands for ${\cal N}\fin(< R\h, >\clM)$, where $R\h$ is the total halo radius, and ${\cal N}\fin(< r)$ stands for ${\cal N}\fin(< r, >0)$ (or for ${\cal N}\fin(< r, < M\h)$, where $M\h$ is the total halo mass). Lastly, the differential form with respect to any argument, $r$ or $\clM$, of any of the previous functions is denoted without the corresponding preceding inequality symbol. For example, ${\cal N}\fin(r,\clM)$ stands for the double derivative with respect to $r$ and $\clM$ and ${\cal N}\fin(\clM)$ stands for the differential subhalo mass function. This greatly simplifies the notation as it avoids writing the dumb arguments $R\h$ or $M\h$ (or 0) as well as the symbols of single and multiple derivatives in most expressions.

Also like in Papers I and II, $M\h$ is defined as the mass encompassed by the virial radius $R\h$ within which the inner mean density is equal to the virial overdensity \citep{BN98,H00} times the current mean cosmic density. In particular, we assume the MW mass equal to $M\h=2.2\times 10^{12}$ \modotc. The cosmology adopted is that given by the best {\it WMAP7} parameters \citep{Km11}, with CDM spectrum according to the prescription given by \citet{BBKS} with the \citet{S95} shape parameter. The reader is referred to Papers I and II for the role of diffuse dark matter (dDM) in the properties of substructure, just briefly referred to in this Paper.

\section{Purely Accreting haloes}\label{PA}

During accretion haloes evolve inside-out because the later particles (and subhaloes) fall onto them, the larger their initial turn-around radius as well as their final apocentric radius due to the ordered virialisation process taking place in this case \citep{SM19}. On the contrary, the violent relaxation suffered by haloes in major mergers causes them to lose the memory of their past history, so that their final properties are indistinguishable from those of purely accreting haloes with the same mass $M\h$ at the same cosmic time $t\h$. \citet{SM19} provide a formal proof for this important result, but the origin of it is as follows. There is a one-to-one correspondence between haloes with mass $M\h$ at the cosmic time $t\h$ and their seeds: peaks of density contrast $\delta$ in the initial Gaussian random density field filtered with a Gaussian window of scale $S$. That correspondence does not involve any other halo or peak characteristic. In particular, it does not depend on how clumpy the initial mass distribution is inside the initial patch encompassed by the filter or, equivalently, on how lumpy the collapse of the peak is. 

Therefore, all properties of haloes with $M\h$ at $t\h$ arising from gravitational clustering, i.e. from their mass assembly, through accretion and major mergers, are degenerate with respect to their formation time. This is why to study them one has the right to assume pure accretion. In fact, since gravitation is scale-free, such halo properties would be strictly universal (i.e. independent of halo mass and formation time), except for the length scale introduced by the power-spectrum of density fluctuations in the CDM cosmology, at the base of the typical mass-concentration $M$--$c$ relation (see \citealt{Jea19} in progress).  

However, subhalo stripping is not related to gravitational clustering and its effects on subhaloes are not erased by violent relaxation. Consequently, stripped subhaloes do retain the memory of the halo assembly history. This is why {\it substructure is expected to depend on halo formation time} and, since haloes with different masses have different typical formation times, {\it on halo mass as well}. In this sense, the properties of substructure derived in Paper II for purely accreting haloes might substantially differ from the properties of ordinary haloes having suffered major mergers. Nevertheless, to understand the latter we need first to comprehend the former from which they follow (see Sec.~\ref{ordinary}). It is thus worthwhile reminding the results of Paper II.

As explained in Papers I and II, we distinguish between ``the time of accretion of a subhalo" onto the halo and ``the time of its first crossing". The latter corresponds to the first time the subhalo orbits within the (virialised and non-virialised parts of) the halo after reaching turnaround. During the first few crossings of the system, subhalo orbits shrink due to their energy exchange with the shells they cross, which causes the non-virialised part of the subhalo to contract adiabatically. But, after these few initial crossings and neglecting the effects of dynamical friction, subhalo orbits stabilise, with the apocentre at the instantaneous virial radius of the (newly virialised part of the) halo, which thus grows inside-out. The time at which subhalo orbits become stable is what we adopt as the time of their accretion onto the virialised halo.\footnote{Strictly speaking, the continuous arrival of new subhaloes that cross the virialised halo causes it a slight abiabatic contraction. But the characteristic time scale of this effect is very long and it can be safely ignored. This is in fact the reason why we can see the inner halo as virialised despite the continuous non-stabilised recent arrivals crossing it.} In what follows we concentrate in monitoring the stripping of subhaloes after their accretion onto the host halo, i.e. once their orbits are fixed (neglecting dynamical friction). 

The stripped subhalo abundance per infinitesimal truncated mass and radius at $\clM\tr$ and $r$ within a purely accreting halo with $M\h$ at $t\h$ is given by 
\beqa 
{\cal N}\fin(r,\clM\tr)={\cal
  N}\tr(r,\clM\tr)~~~~~~~~~~~~~~~~~~~~~~~~~~~~~~~~~~~~\nonumber\\ 
+\Bigg\lav\!\!\int_{\clM}^{M(r)}\!\!\der M\,{\cal N}\acc(r,M) \!\!\int_{R\tr(v,r,M)}^{R(r,M)}\!\!\!\! \der r'{\cal
    N}\fin_{\rm [M,t(r)]}(r',\clM\tr)\!\!\Bigg\rav,\!\!\!\!
\label{corrbis} 
\eeqa
where angular brackets indicate average over the tangential velocity $v$ of subhaloes at their apocentre at $r$ where they spend most of the time. The subindex [M,t] in the properties of subhaloes is to indicate that they refer to host halos with $M$ at $t$. For simplicity in the notation, we have skipped the subindex [M$\h$,t$\h$] for the host halo itself, but we will re-introduce it in Sections \ref{ordinary} and \ref{specific} when dealing with haloes of different masses and times.  

The first term on the right of equation (\ref{corrbis}), equal to
\beq
{\cal N}\tr(r,\clM\tr)
= \mu(r,\clM\tr)\, {\cal N}\acc(r,\clM\tr),
\label{firstterm} 
\eeq
gives the contribution directly arising from the stripping of accreted subhaloes of suited mass $\clM$, which have their apocentre at $r$ (see Paper I) and whose abundance is 
\beq
{\cal N}\acc(r,\clM)\!=\!4\pi r^2 \frac{\rho(r)}{M\h}\,{\cal N}\acc(\clM),
\label{nacc}
\eeq
where ${\cal N}\acc(\clM)$ is their differential MF and $\rho(r)$ is the halo density profile. In equation (\ref{firstterm}), $\mu(r,\clM\tr)$ is the mean (averaged over $v$) truncated-to-original mass ratio of subhaloes with $\clM\tr$ at $r$, calculated in Paper II by monitoring the mass loss through repetitive stripping and shock heating of subhaloes accreted at $t(r)$ when the host halo had radius $r$ and mass $M(r)$. And the second term on the right of equation (\ref{corrbis}) gives the contribution arising from subsubhaloes that were lying in accreted subhaloes with mass $M$ and tangential velocities $v$ at $r$ and have been released into the intra-halo medium when their hosts, with initial radius $R(r,M)$, have been truncated at $R\tr(v,r,M)$. See Paper II for the expression of the truncation radius in objects with the NFW \citep{NFW97} density profile.

\begin{figure}
\centerline{\includegraphics[scale=.45,bb= 18 60 560 558]{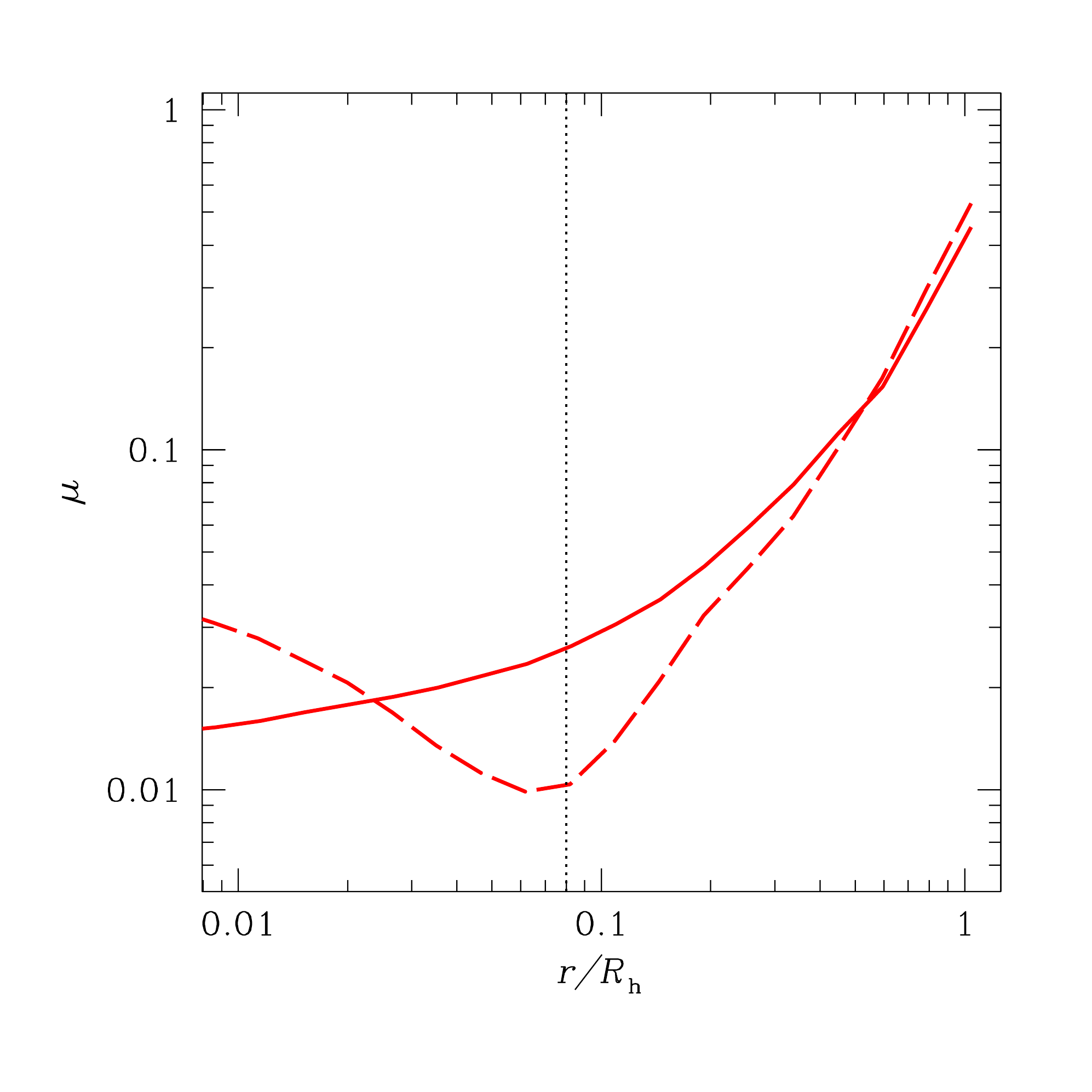}}
\caption{Mean truncated-to-original subhalo mass ratio profile predicted by CUSP for subhaloes with $\clM=10^{8}$ \modot in purely accreting MW-mass haloes using the unbiased CUSP $M$--$c$ relation (solid red line) and the \citet{Gea08} empirical $M$--$c$ relation affected by the limited mass resolution of simulations (long-dashed red line). The vertical dotted black line marks the radius where the inwards decreasing behaviour of $\mu$ in the latter case is inverted for the reason explained in the text.}
(A colour version of this Figure is available in the online journal.)
\label{f1}
\end{figure}

Equation (\ref{corrbis}) can be rewritten in the simple form 
\beq
{\cal N}\fin(r,\clM\tr)=[1+f\rel(r,\clM\tr)]\,\mu(r,\clM\tr)\,{\cal N}\acc(r,\clM\tr),
\label{M} 
\eeq
where $f\rel(r,\clM\tr)$ is the virtual fraction of accreted subhaloes with mass $\clM\tr$ converted into stripped ones of that mass at $r$ arising from subsubhaloes. This fraction is the solution of the Fredholm integral equation of the second kind (with the boundary condition $f\trb(0,\clM\tr)=0$) of the differential equation
\beq 
-\frac{\der }{\der r}\left[f\trb(r,\clM\tr)\,\mu(r,\clM\tr)\frac{R\h^3}{r^3}\right]
\!=\!\Bigg\lav
\frac{\partial {\cal R}}{\partial r}
\frac{{\cal N}\fin({\cal R},\clM\tr)}{{\cal N}\acc(\clM\tr)}\Bigg\rav ,
\label{final3bis}
\eeq
with ${\cal R}(v,r)\equiv R\tr(v,r,\clM) R\h/\clR(r,\clM)$. The function $f\rel$ is always less than a few percent, so it can be safely neglected in front of unity, though, for the sake of completeness, it is kept in all the expressions below. On the contrary, the mean truncated-to-original subhalo mass ratio, $\mu(r,\clM\tr)$, depicted in Figure \ref{f1} plays a crucial role in the properties of substructure. It is thus worth explaining its main features. 

As the strength of stripping and shock heating depends on the concentration $c$ of both subhaloes and the host halo (see Paper II) and $c$ depends on the mass of the object through the well-known mass-concentration ($M$--$c$) relation, $\mu$ depends on the masses of subhaloes and the host halo. Although those dependencies are not gathered in Figure \ref{f1}, which focuses on the $\mu$ profile for subhaloes with $10^8$ \modot in haloes with $2.2 \times 10^{12}$ \modotc, they are very important to understand the properties of substructure. Indeed, the $\mu$ varies in a non-trivial way according to the mass of the host halo: at $R\h$ it is always close to unity, but its inwards decrease is less steep in more massive haloes (see Fig.~\ref{f5}) because their concentration is lower, which causes the pericentric radius reached by subhaloes with identical $v$ at apocentre to be larger so that stripping and shock heating is less marked (see Paper II). As a result, the mass integral of $\mu$ varies with halo mass as $\propto M\h^{0.08}$. Regarding the dependence of $\mu$ on subhalo mass, it turns out that $\mu(r,\clM\tr)$ is separable. The reason for this is that subhaloes are truncated by tidal stripping at the radius (dependent on their own concentration) where the inner mean density equals that of the halo at the pericentre, which is the same for subhaloes of all masses with identical $v$ at $r$ (see Paper II). The factor dependent on subhalo mass is essentially proportional to $({\clM\tr})^{-0.03}$. Both mass dependencies are weak, however, particularly that on subhalo mass due to the fact that subhaloes accreted at any time $t(r)$ have similar concentrations. In this sense the $\mu$ profile for any fixed halo mass can be seen, in a first approximation, to depend only on $r$. 

A more subtle issue is that, for the above mentioned reasons, the shape of the $\mu$ profile will depend on the $M$--$c$ relation. Of course, the real $\mu$ profile predicted by CUSP implicitly follows from the $M$--$c$ relation that can be derived within that framework \citep{Jea19}. However, if we want to reproduce the results of simulations, we must use the $M$--$c$ relation found in numerical studies by e.g. \citet{Gea08} with a limited mass resolution similar to that affecting those empirical results. The effect of the limited mass resolution of simulations, which affects all (sub)haloes at early times when they are little massive, is apparent in Figure \ref{f1}. While the $\mu$ profile derived from the CUSP $M$--$c$ relation with no limited mass resolution is ever decreasing inwards, that found with the \citet{Gea08} $M$--$c$ relation stops decreasing at $r\sim 0.08R\h$ and then begins to increase again. The reason for this strange result is the following. In a purely accreting halo evolving inside-out as considered here, its concentration at earlier times decreases with decreasing $r$ as $r/\rs$, because of the fixed value of $\rs$. Since the concentration of accreted subhaloes also decreases with increasing $z$ (this is so in all empirical as well as theoretical $M$--$c$ relations), stripping keeps on being effective at small radii populated by subhaloes accreted at higher redshifts. However, due to the limited mass resolution, the concentration of (sub)haloes in the \citet{Gea08} $M$--$c$ relation is bounded to a minimum value independent of mass reached at$z \sim 3$ ($r\sim 0.3 R\h$). As a consequence, at radii smaller than $0.3 R\h$, the concentration of the inside-out evolving host halo continues to decrease, while that of subhaloes does not, and subhalo stripping becomes (artificially) ineffective.

The previous discussion also illustrates that, in normal conditions, the $\mu$ profile decreases inwards despite the fact that the smaller the radius, the lower the concentration of the host halo seen by subhaloes. Indeed, the main cause shaping the $\mu$ profile is the time subhaloes at different radii have been undergoing stripping rather than the different typical tidal forces they see. 

\begin{figure}
\centerline{\includegraphics[scale=.45,bb= 18 60 560 558]{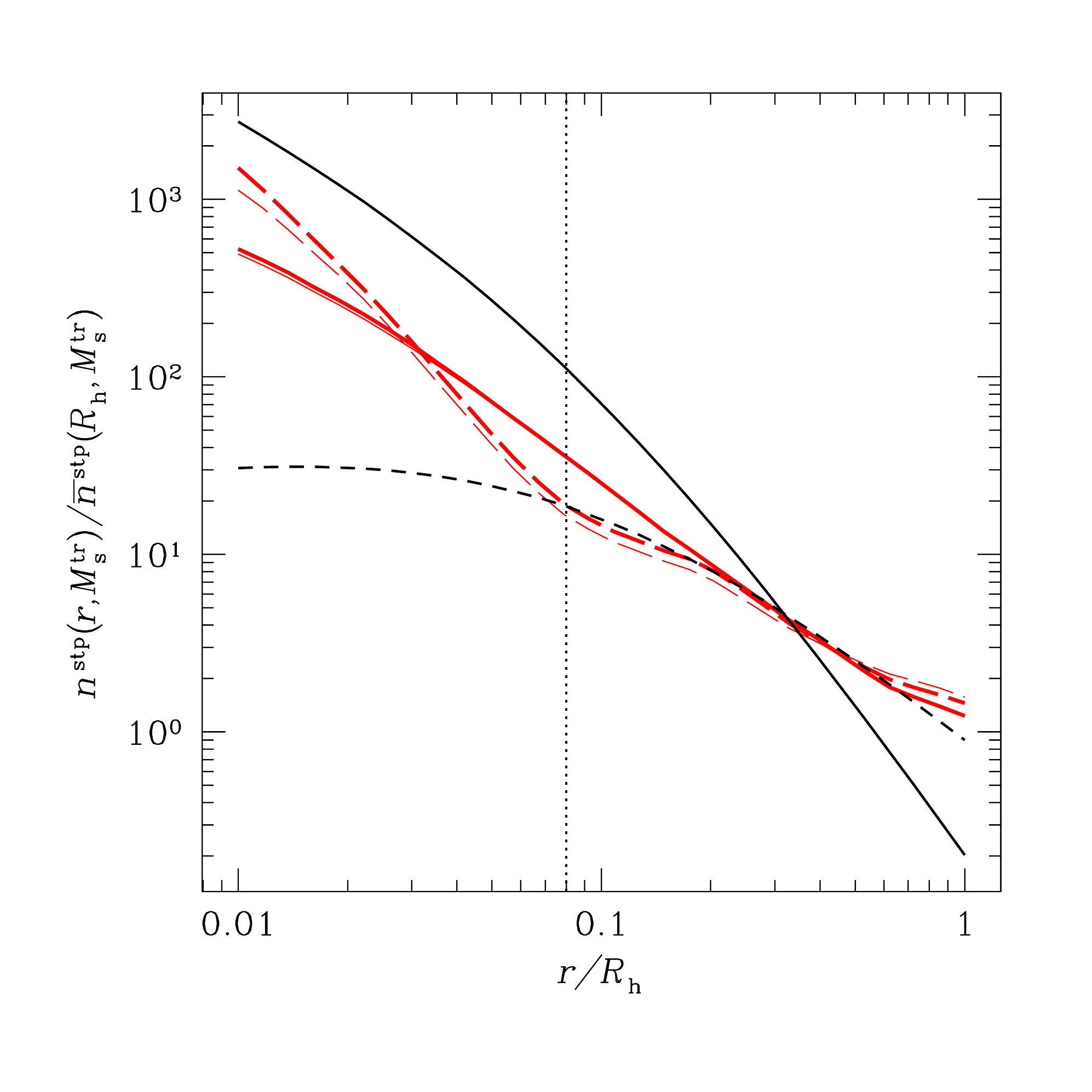}}
\caption{Scaled number density profiles of subhaloes with different masses (red lines) predicted for purely accreting haloes with current MW-mass using the CUSP $M$--$c$ relation (solid red lines) and the \citet{Gea08} $M$--$c$ relation affected by the limited mass resolution of simulations (long-dashed red lines). For comparison we plot the fit by \citet{Han15} to the profile found in the halo A of the Level 1 Aquarius simulation (black dashed line) affected by a similar resolution but having not evolved by accretion before $z\sim 6$ corresponding to the radius marked with a vertical dotted black line like in Fig. \ref{f1}. The solid black line is the scaled halo density profile. To better appreciate the effect of changing the subhalo masses, we plot the predictions for subhaloes with $10^{-2}M\h$ and $10^{-4}M\h$ in thick and thin red lines, respectively.}
(A colour version of this Figure is available in the online journal.)
\label{f2}
\end{figure}

\begin{figure}
\includegraphics[scale=.45,bb= 18 50 560 558]{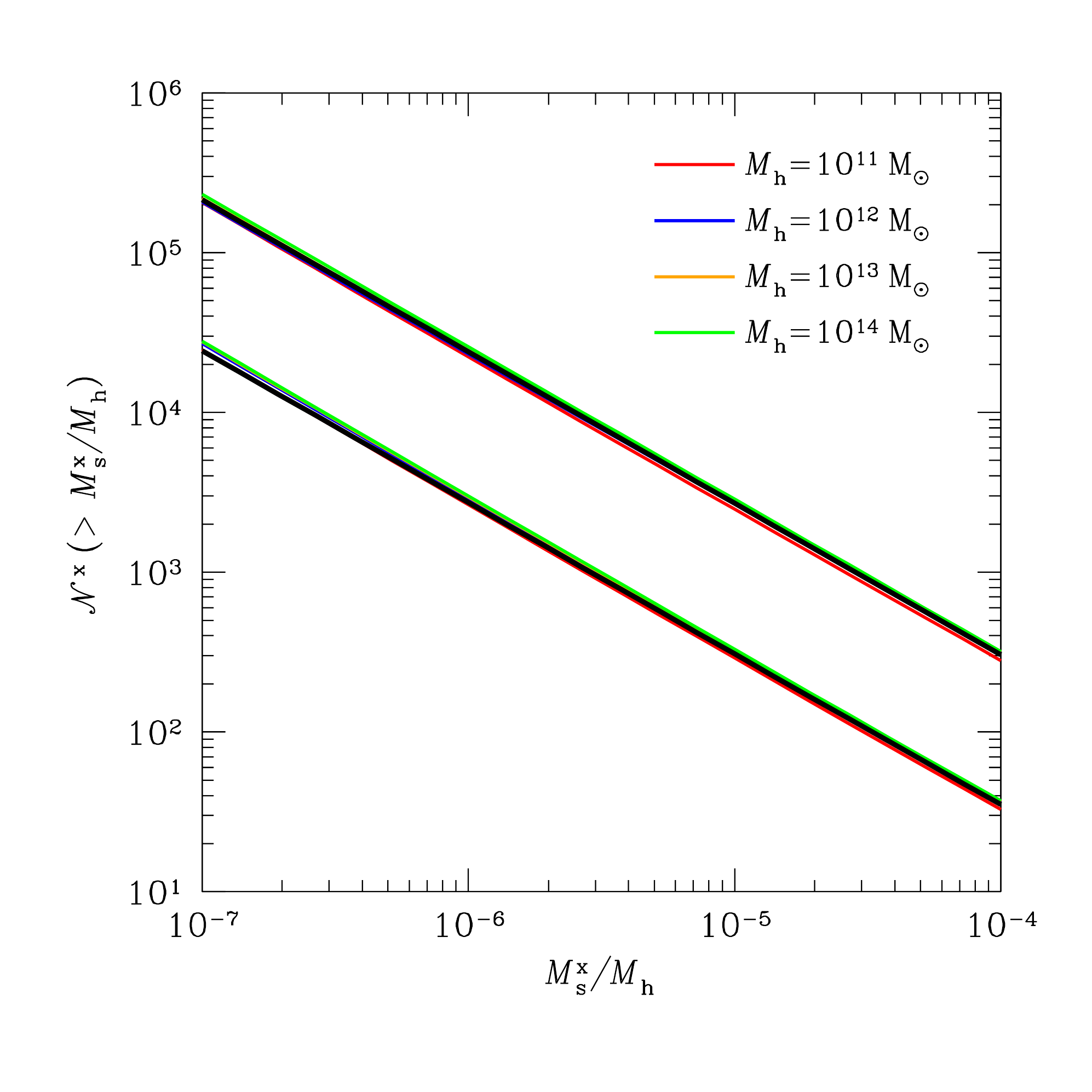}
 \caption{Cumulative MF of {\it accreted} subhaloes (upper curves) and of {\it stripped} subhaloes (lower curves) (superindex ``x" on ${\cal N}$ stands for ``acc" or ``str”, respectively) predicted by CUSP for purely accreting haloes of the different quoted masses $M\h$. At the scale of the plot, the predictions for the Gao et al. or CUSP $M$--$c$ relations coincide. The MFs of accreted subhaloes for haloes with different masses overlap as their counterparts in simulated ordinary haloes (solid black lines) and the same is true for the MFs of stripped subhaloes provided they are multiplied by $[M\h/(10^{12}$ \modotc$)]^{-0.08}$. These results agree with what is found in simulated ordinary haloes (solid black lines) as derived by \citealt{Hea18}, here properly normalised so as to include first-level subhaloes only (see Paper II).}
 (A colour version of this Figure is available in the online journal.)
\label{f3}
\end{figure}

Since both $\mu(r,\clM\tr)$ and ${\cal N}\acc(r,\clM\tr)$ are separable, so is also the radial abundance of stripped subhaloes ${\cal N}\fin(r,\clM\tr)$ (eq.~[\ref{M}]). Thus, the number density profile per infinitesimal mass of stripped subhaloes, $n\fin(r,\clM\tr)\equiv {\cal N}\fin(r,\clM\tr)/(4\pi r^2)$, scaled to their total number in the halo,
\beq
\frac{n\fin(r,\clM\tr)}{\bar n\fin(R\h,\clM\tr)}\!=\!
\frac{{\cal N}\fin(r,\clM\tr)/(4\pi r^2)}{3{\cal N}\fin(\clM\tr)/(4\pi R\h^3)},
\label{new}
\eeq
is independent of subhalo mass. From now on a bar on a quantity dependent on $r$ means the radial average of that quantity inside $r$. 

In Figure \ref{f2} we show the scaled number density profiles of subhaloes of two masses ($10^{-2}M\h$ and $10^{-4}M\h$) found for the two above mentioned $M$--$c$ relations. For comparison we plot the profile found by \citet{Han15} in the MW-mass halo A of the Level 1 {\it Aquarius} simulation \citep{Sea08}, also found to be roughly independent of subhalo mass (except for the effects of dynamical friction; \citealt{Hea18}). This halo is particularly well suited to the comparison with our predictions for purely accreting haloes because it suffered the last major merger at $z\sim 6$ ($r\sim 0.08 R\h$) and has been evolving by accretion (and growing inside-out) since then. Note that, given its formation time, the concentration $r/\rs$ cannot be traced down to radii smaller than $r=0.08R\h$. This is the reason that its $\mu$ profile is not seen to increase at smaller radii. In fact, as we will see in Section \ref{ordinary}, it should be flat there, though the $\mu$ profile of the simulated halo is not well determined at those radii because it is dominated by orphan objects.\footnote{Accreted subhaloes are dubbed ``orphan" when their stripped subhaloes have masses below the mass resolution of the simulation.}

Like Figure \ref{f1}, Figure \ref{f2} focuses on MW-mass haloes so it does not inform on the dependence on halo mass of the scaled subhalo number density profile in purely accreting haloes. However, our calculations show that the profiles for haloes of different masses are quite similar. They do not overlap, however, because, as mentioned, the corresponding $\mu$ profiles are not simply shifted with respect to each other by a constant factor (see Fig.~\ref{f6} dealing with ordinary haloes).

Lastly, integrating over $r$ the subhalo abundance given in equation (\ref{M}), we obtain the subhalo differential MF, which takes the form
\beq
{\cal N}\fin(\clM\tr)=\overline {(1+f\trb)\mu}(R\h)\,{\cal N}\acc(\clM\tr)\,.
\label{MF}
\eeq
And integrating it over subhalo mass from $\clM\tr$, we arrive at the cumulative MF, ${\cal N}\fin(>\clM\tr)$. 

In Figure \ref{f3} we plot the cumulative MFs of accreted and stripped subhaloes predicted for purely accreting haloes. As can be seen, the MF of accreted subhaloes is universal, i.e. independent of halo mass, in agreement with the results of simulations (\citealt{Hea18} and references therein). This is equivalent to say that the differential scaled subhalo abundance ${\cal N}\acc(\clM/M\h)$ per infinitesimal $\clM/M\h$ is also universal or that the subhalo abundance ${\cal N}\acc(\clM/M\h)$ per infinitesimal $\clM$ varies with halo mass as $M\h^{-1}$ (not to mix up with the differential subhalo abundance ${\cal N}\acc(\clM)$ per infinitesimal $\clM$, which is roughly proportional to $M\h$; e.g. \citealt{Fea10}). The subhalo abundance ${\cal N}\acc(\clM/M\h)$ per infinitesinal $\clM$ which will have important consequences in Section \ref{specific}. In Figure \ref{f3} we also see that the scaled MF of stripped subhaloes is also universal {\it provided it is multiplied by} $M\h^{-0.08}$. Such a dependence on halo mass of the MF of stripped subhaloes following from the above mentioned dependence of the integral over $r$ of their $\mu$ profile as a consequence of the mass dependence of halo concentration fully agrees with that observed in simulations (the factor rendering the MF of simulated haloes universal is $M\h^{\eta}$ with $\eta=-0.1$; \citealt{Hea18,Rea16}; see also \citealt{Zeabis05,Gi08,Gea11}). Strictly speaking, the subhalo MF found in simulations is influenced at the high-mass end by the effects of dynamical friction, ignored in our model. But this effect is insignificant for subhaloes less massive than $10^{-4}M\h$ as represented here (see Fig.~10 in Paper II). There is one caveat, however, in this agreement: while our predictions are for purely accreting haloes, the results of simulations refer to {\it ordinary haloes having suffered major mergers}. In other words, it is not clear whether the agreement is accidental or it will persist when dealing with ordinary haloes. 

The empirical result that less massive haloes are poorer (in the sense that their subhalo MF are lower) than more massive ones is commonly interpreted as due to the different formation times of haloes of different masses. Indeed, the less massive a halo, the earlier it typically forms, so: 1) stripping has more time to proceed and 2) it was more efficient because haloes are denser at high redshifts. However, this explanation is at odds with the recent finding by \citet{Bea20} that the earlier MW-mass haloes form, the richer they are. On the other hand, it is not supported either by the fact that purely accreting haloes show the same dependence despite they all have the same arbitrarily small formation time. Our results rather point to the fact that such a dependence is due to the mass dependence of halo concentration, though we must first confirm that the mass dependence of the MF in purely accreting haloes is preserved in ordinary ones. In fact, our predictions also show that the longer subhaloes have been stripped, the lower their final $\mu$ profile. In other words, the different typical concentrations and typical formation times of haloes of different masses go in the opposite direction, so we must clarify which is the dominant effect and why in the richness of ordinary haloes.

\section{Ordinary Haloes of Different Masses}\label{ordinary}

The density, accreted dDM mass fraction and mean abundance of accreted subhaloes profiles, $\rho(r)$, $f\dc\acc(r)$ and ${\cal N}\acc(r,\clM)$, respectively, are not related to tidal stripping,\footnote{Stripping redistributes the dDM lost by subhaloes in a leading arm and a trailing tail over their orbits. But, for the same reason that the contribution from subhaloes to the halo density profile can be calculated assuming they lie at their apocentre where they spend most of the time, the contribution from the stripped dDM can also be calculated assuming it is located at the apocentre of their orbit. Consequently, stripping does not essentially alter the halo density profile.} so they are the same in both purely accreting and ordinary haloes having suffer major mergers (\citealt{SM19}). Thus, the only functions in the expression of the radial distribution of stripped subhaloes (eq.~[\ref{M}]) that depend on stripping are $\mu$ and $f\rel$ defined in equations (\ref{firstterm}) and (\ref{final3bis}), respectively. Therefore, to obtain the radial distribution of stripped subhaloes in ordinary haloes we must first determine these two functions in such haloes. 

Next we show how to obtain these properties in ordinary haloes from their counterparts in purely accreting ones derived in Section \ref{PA}, hereafter distinguished with superindex PA. We will also use from now the scaled arguments $x=r/R\h$ and $m=\clM/M\h$ so that all functions of those arguments should essentially coincide for purely accreting haloes of different masses. Note that the properties of ordinary haloes of any given mass derived next are their average over all formation times of such haloes. Of course, those theoretical properties are hard to compare to the results of simulations which at present provide the properties of substructure in a small number of haloes of all masses. Nevertheless, these mean properties allow us to elucidate the origin of their dependence on halo mass without being disturbed by the statistical deviations of individual objects.

Be $F_{\rm [M\h,t\h]}(x,m)$ the fraction of accreted subhaloes per infinitesimal mass and radius that satisfy some condition in haloes of $M\h$ at $t\h$ averaged over their formation time (i.e. the time of their last major merger) and $F\PA_{\rm [M\h,t\h]}(x,m)$ its counterpart in purely accreting haloes. Taking into account the inside-out growth of haloes after their last major merger, we have the following relation between the two quantities
\beqa
F_{\rm [M\h,t\h]}(x,m)=\int_0^{t(x)} \der t \,f_{\rm [M\h,t\h]}(t)\,F\PA_{\rm [M\h,t\h]}(x,m)~~~~~~~\nonumber\\
+  \int_{t(x)}^{t\h} \der t\, f_{\rm [M\h,t\h]}(t)\, 
\bar F_{\rm [M(t),t]}(1,m),~~~~~~~~~~
\label{one2}
\eeqa
where $f_{\rm [M\h,t\h]}(t)$ is the formation time probability distribution function (PDF) of haloes with $M\h$ at $t\h$, calculated within the CUSP formalism in \citet{Mea98} (see also \citealt{Rea01} for a practical approximate expression in the extended Press-Schechter (EPS) formalism; \citealt{PS,B91,BCEK,LC94}). $\bar F_{\rm [M(t),t]}(x,m)]$ stands for the mean fraction $F_{\rm [M(t),t]}(x,m)]$ inside $x$ of accreted subhaloes, but, given the form of ${\cal N}\acc(x,m)$ (eq.~[\ref{nacc}]), it coincides with the simple radial average of $F_{\rm [M(t),t]}(x,m)$ inside $x$. This is why we denote it with a bar. Note that the contribution on $F(x)$ from haloes formed after $t(x)$ (the second term on the right) is averaged over their own formation times. It thus takes into account the different weight of haloes formed at $t(x)$ with previous different formations times (and so on so forth). On the contrary, the contribution on $F(x)$ from haloes formed before $t(x)$ (the first term) does not depend on their individual formation times ($F\PA$ does not depend on any formation time) because, at the radius $x$, such haloes are accreting material ex-novo.

\begin{figure}
\centerline{\includegraphics[scale=.45,bb= 18 50 560 558]{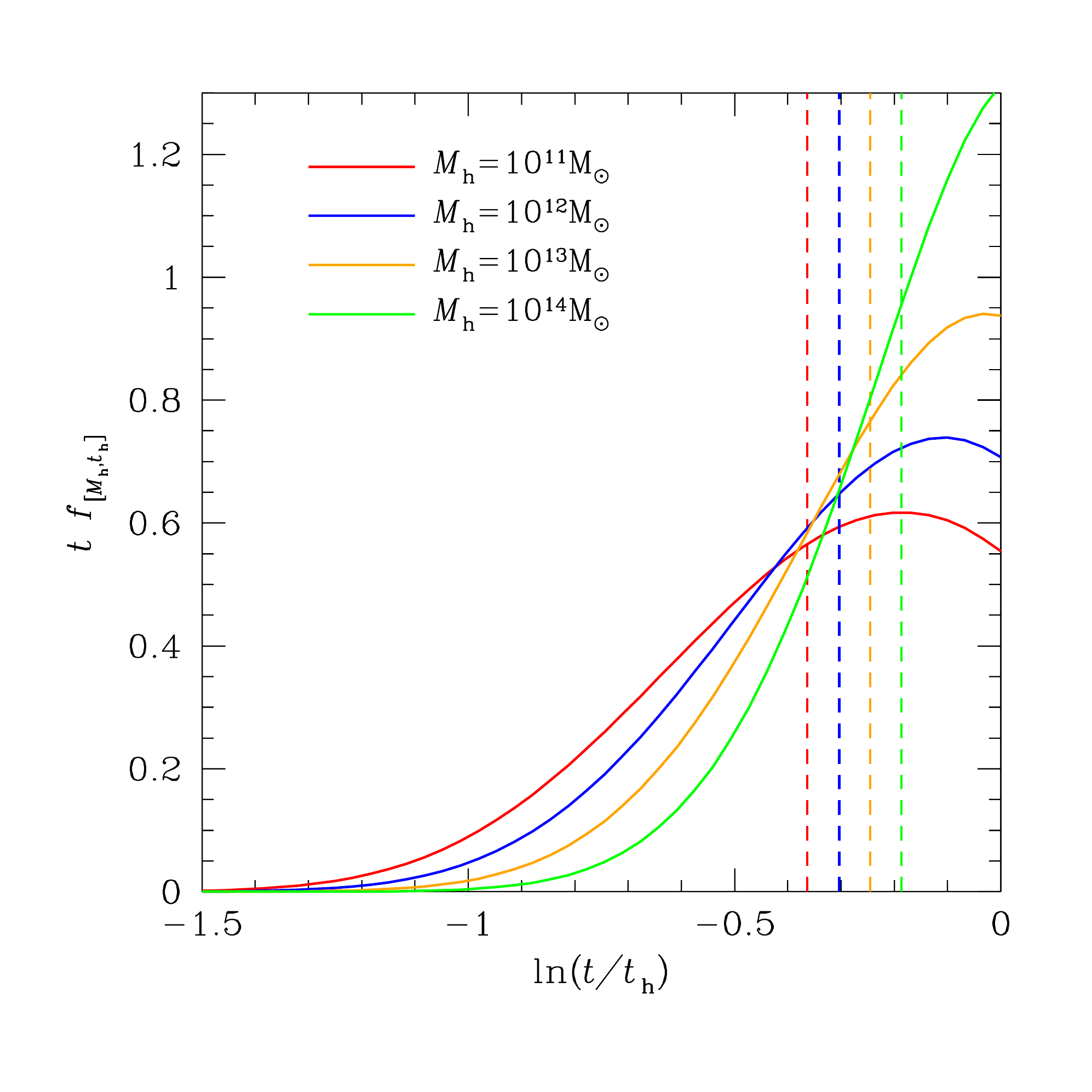}}
\caption{Formation time PDFs of haloes with the quoted masses (coloured lines) at the present time $t\h=t_0$. The coloured dashed vertical lines mark the median formation time $t\med$ of haloes of each mass.}
(A colour version of this Figure is available in the online journal.)
\label{f4}
\end{figure}

To write equation (\ref{one2}) we have taken into account that, when a halo suffers a major merger, its content is scrambled,\footnote{The scrambling must be complete, otherwise major mergers would not cause haloes to fully lose the memory of their past history as they do \citep{SM19}.} so the fraction $F_{\rm [M(t),t]}(x,m)$ at any radius $x$ equals its mean value within the total radius at that moment, $\bar F_{\rm [M(t),t]}(1,m)$. In addition, we have taken into account that, after the last major merger, haloes evolve by pure accretion, so $M(t)/M\h$ is the mass track of purely accreting haloes with boundary condition $M(t\h)/M\h=1$. Note that, even if $F\PA_{\rm [M\h,t\h]}(x,m)$ (in scaled arguments) is essentially universal, $F_{\rm [M\h,t\h]}(x,m)$ will depend on $M\h$ and $t\h$ through the explicit dependence on these quantities of the halo formation time PDF (Fig.~\ref{f4}).  

Multiplying equation (\ref{one2}) by ${\cal N}\acc(x,m)$ and integrating over $x$ out to 1, we are led, by partial integration and taking into account equation (\ref{nacc}), to
\beqa
\bar F_{\rm [M(t\h),t\h]}(1,m)=\overline {F\PA_{\rm [M\h,t\h]}\,f_{\rm [M\h,t\h]}\uc}(1,m)~~~~~~~~~~~~~~~~~~~\nonumber\\
\!+\!\!\int_0^{t\h}\!\! \der t\, f_{\rm [M\h,t\h]}(t)
\bar F_{\rm [M(t),t]}(1,m)\frac{M(t)}{M\h},~~~~~~~~~~~~
\label{one3}
\eeqa
where $f_{\rm [M\h,t\h]}\uc(x)$ stands for the cumulative formation time PDF of haloes with $M\h$ at $t\h$ up to $t(x)$. Equation (\ref{one3}) is a Volterra equation of second kind for $\bar F_{\rm [M(t),t]}(1,m)$ as a function of $t$. Note that, according to equation (\ref{one3}), the mass average of $F$ out to 1, $\bar F_{\rm [M(t),t]}(1,m)$, is different from $\bar F\PA{\rm [M(t),t]}(1,m)$. The reason for this will be seen below. Bringing the solution of this Volterra equation in the integral on the right of equation (\ref{one2}), we arrive at the desired function $F_{\rm [M(t\h),t\h]}(x,m)$ for any value of $x$ (and $m$). Note also that, as $n_{\rm [M\h,t\h]}\uc(x)$ does not depend on $m$, equations (\ref{one3}) and (\ref{one2}) imply that $F_{\rm [M(t\h),t\h]}(x,m)$ would be independent of $m$ provided its counterpart $F\PA$ were. (This is approximately the case for the $\mu$ and $f\rel$ functions; see next.)

\begin{figure}
\centerline{\includegraphics[scale=.45,bb= 18 50 560 558]{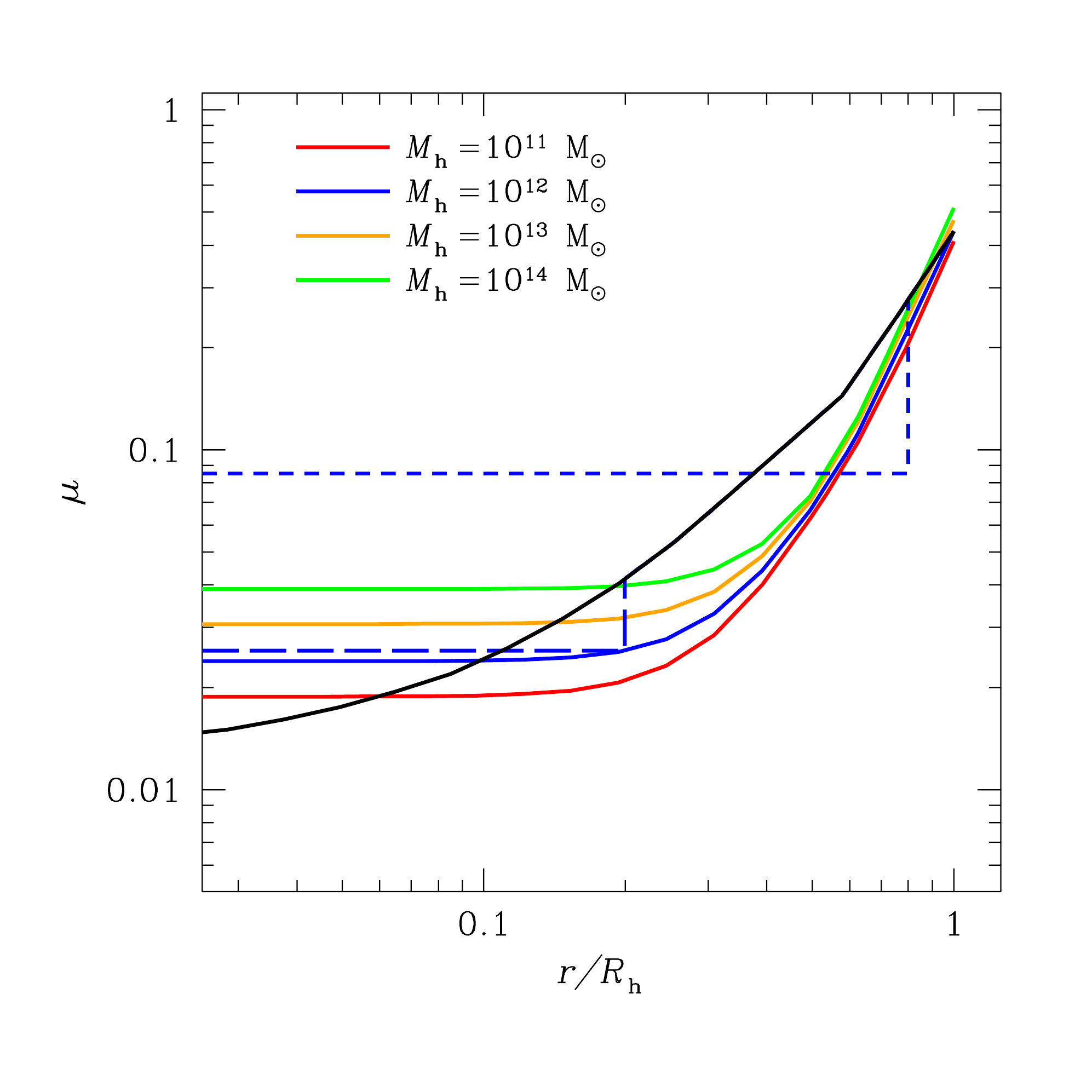}}
\caption{Mean truncated-to-original subhalo mass ratio profiles predicted using the CUSP $M$--$c$ relation for ordinary haloes of several masses $M\h$ at the current time $t_h$ averaged over their formation times (solid coloured lines). The results for different subhalo masses $\clM\tr$ (from $10^6$ \modot to $10^9$ \modotc) overlap when they are multiplied by $[\clM\tr/10^8$ \modotc$)]^{-0.03}$. For comparison we plot the prediction for purely accreting MW-mass haloes (solid black line). To illustrate the effects of averaging over halo formation times we also plot in long-dashed and short-dashed lines the profiles for two {\it individual} haloes of $10^{12}$ \modot formed at a high- and low-redshift, respectively.}
(A colour version of this Figure is available in the online journal.)
\label{f5}
\end{figure}

This procedure can be applied to the virtual fractions of accreted subhaloes with $\clM\tr$ at $r$ converted into stripped subhaloes by direct stripping and through the release of subsubhaloes we arrive at the $\mu(x,m)$ and $f\rel(x,m)$ profiles in ordinary haloes of any mass averaged over their formation times. In Figure \ref{f5} we depict the $\mu(x,m)$ profile of ordinary haloes averaged over their formation times that is predicted for the CUSP $M$--$c$ relation. To better realise the effect of that average, we also plot the schematic (approximate) $\mu$ profiles of two individual haloes of $10^{12}$ \modotc, one formed at $t\f\sim 2$ Gyr (corresponding to $r=0.2 R_h$ in the final halo grown inside-out since that moment) and the other one formed at $t\f\sim 10$ Gyr (corresponding to $r=0.8 R\h$). For the reason mentioned above when explaining the meaning of $\bar F_{\rm [M(t),t]}(1,m)$ in the second integral on the right of equation (\ref{one2}), the scrambling of the system at the last major merger yields a flat $\mu$ profile equal to its radial average within the formation radius. This is not only the case just after the merger, but also long time after. Indeed, after the major merger all subhaloes inside the scrambled region orbit and suffer stripping during the same time interval $t\h-t\f$ regardless of their past history, so $\mu$ in that region deepens, but it remain essentially flat until the next merger or the final time. Strictly speaking, subhaloes at different radii within that region suffer different tidal forces over their orbits (the smaller the radius, the lower the concentration of the halo they see). But, as mentioned in connection with Figure \ref{f1}, this has a much less marked effect on $\mu$ than the different time subhaloes suffer stripping. In fact, if the ending $\mu$ profile within that region did show a substantial dependence on radius, the accurate $\mu$ profile of haloes averaged over their formation times commented below would show it, which is not the case (see Fig.~\ref{f5}). Likewise, the flat $\mu$ profile of those individual halos inside the formation radius have been taken equal to the radial average inside that radius of the $\mu\PA$ at the ending time, while it is actually somewhat lower due to the stripping suffered by subhaloes during the time elapsed since the scrambling. Besides those simplifications, the important point to retain from those schematic examples is that, since after the merger haloes evolve inside-out by accretion, their (essentially flat) $\mu$ profile in the inner region jumps at its edge to the $\mu\PA$ profile of purely accreting haloes with the same mass $M\h$ at $t\h$.

That behaviour of the $\mu$ profile in individual haloes of a given mass translates into their formation-time average. This is the reason why the formation-time-averaged profile is also flat at small radii and begins to increase at some radius, dependent on the typical formation time of haloes of that mass, towards the profile of purely accreting haloes. The difference is that, while the profiles of individual haloes reach the purely accreting solution right at the formation radius, the formation-time-averaged profile only reaches it at $R\h$ so that it stays systematically below that solution at large radii. The reason for this difference is clear. Since the $\mu$ profiles are outwards increasing, their mass average inside any formation radius is always lower than the original value at that radius, so the formation-time-averaged profile is also lower than the purely accreting solution. And, as large radii contribute the most to the mass average of $\mu$ inside $R\h$, that mass average in ordinary haloes averaged over their formation times will always be somewhat smaller than that of purely accreting haloes. In any event, the difference should be similar for haloes of any mass, which implies that the mass average of $\mu$ in ordinary haloes will essentially coincide with that in purely accreting ones arising, as mentioned in Section \ref{PA}, from the mass dependence of halo concentration. All these conclusions referring to the $\mu$ profiles of ordinary haloes will translate into their subhalo radial distributions and MFs, which will explain their behaviour. 

\begin{figure}
\centerline{\includegraphics[scale=.45,bb= 18 50 560 558]{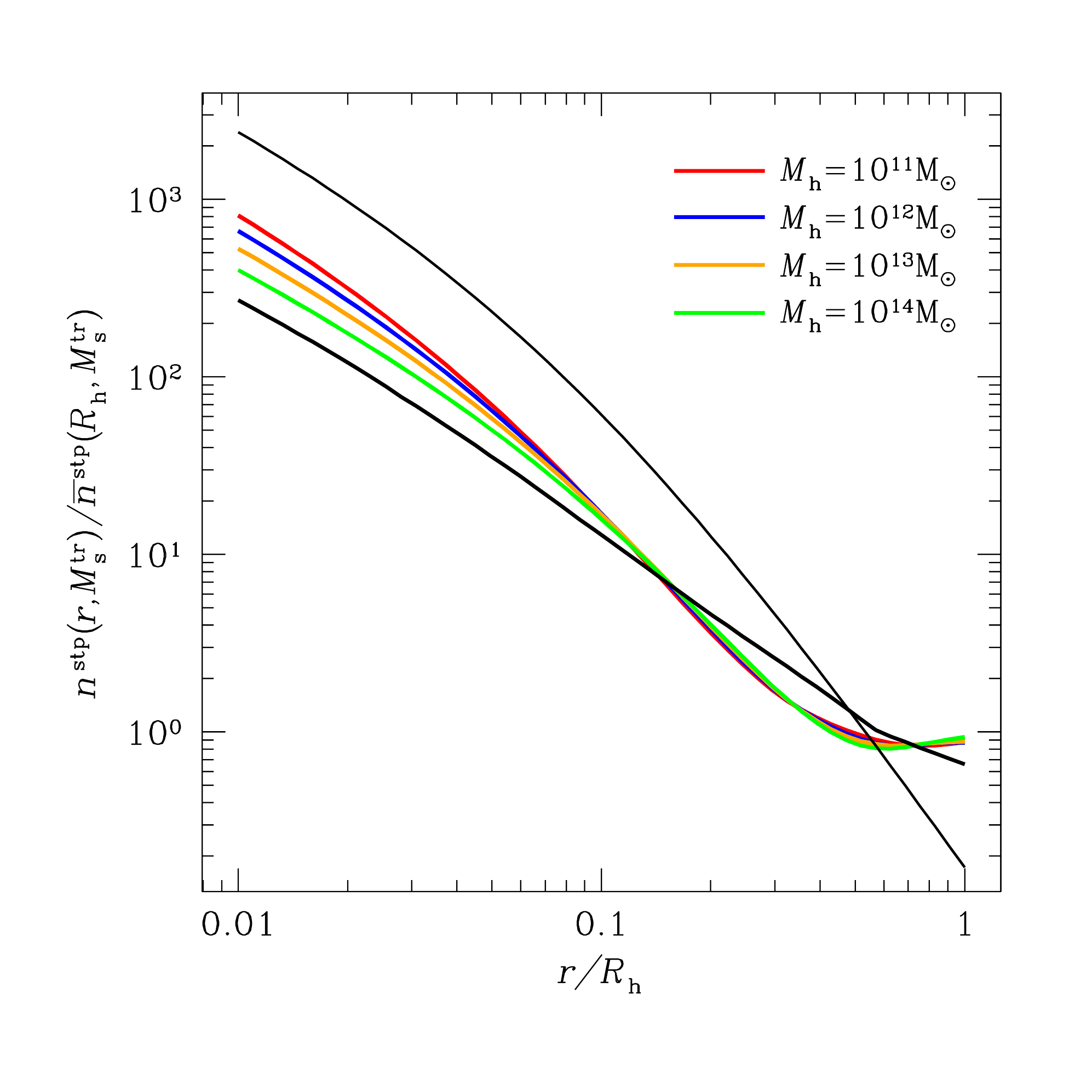}}
\caption{Scaled number density profiles of subhaloes of different masses $\clM$, which overlap with no added factor, in ordinary haloes of several masses $M\h$ (coloured lines), compared to the profile obtained in purely accreting haloes of $10^{12}$ \modot (thick black line) and the scaled mass density profile of such haloes (thin black line).}
(A colour version of this Figure is available in the online journal.)
\label{f6}
\end{figure}

Once the functions $\mu$ and $f\rel$ have been determined, we can proceed to derive the subhalo abundance per infinitesimal mass and radius around $r$ and $\clM\tr$, ${\cal N}\fin(r,\clM\tr)$, in ordinary haloes by application of equation (\ref{M}). Specifically, since the abundance of accreted subhaloes does not depend on the halo formation time and $f\rel$ is negligible, the subhalo radial abundance in ordinary haloes of a given mass averaged over their formation times is simply equal to the formation-time-averaged $\mu$ profile times the abundance of accreted subhaloes. 

The first consequence of the form of the radial abundance of subhaloes in ordinary haloes refers to the scaled subhalo number density profile (eq.~[\ref{new}]). As can be seen in Figure \ref{f6}, the profiles of subhaloes of different masses overlap because their $\mu$ profiles differ, as mentioned, by the same constant factor as in purely accreting haloes, which cancels with their scaling. Also like in purely accreting haloes, the profiles for haloes of different masses do not overlap because their $\mu$ profiles do not differ by just a constant factor. But there is one interesting difference in comparison with the case of purely accreting haloes: the scaled number density profiles are substantially steeper now. In fact, at small enough radii they are parallel to the halo density profile. (To avoid crowding, in Figure \ref{f6} we only plot the mass density profile and the scaled subhalo number density profile for haloes of $10^{12}$ \modotc, so that this comparison is only possible for such haloes.) The reason for such a behaviour of the scaled number density profile of ordinary haloes is that in the scrambled regions subhaloes of all masses have been mixed up with the dDM, so their number density profiles become proportional to the mass density profile of the halo there. 

\begin{figure}
\centerline{\includegraphics[scale=.45,bb= 18 50 560 558]{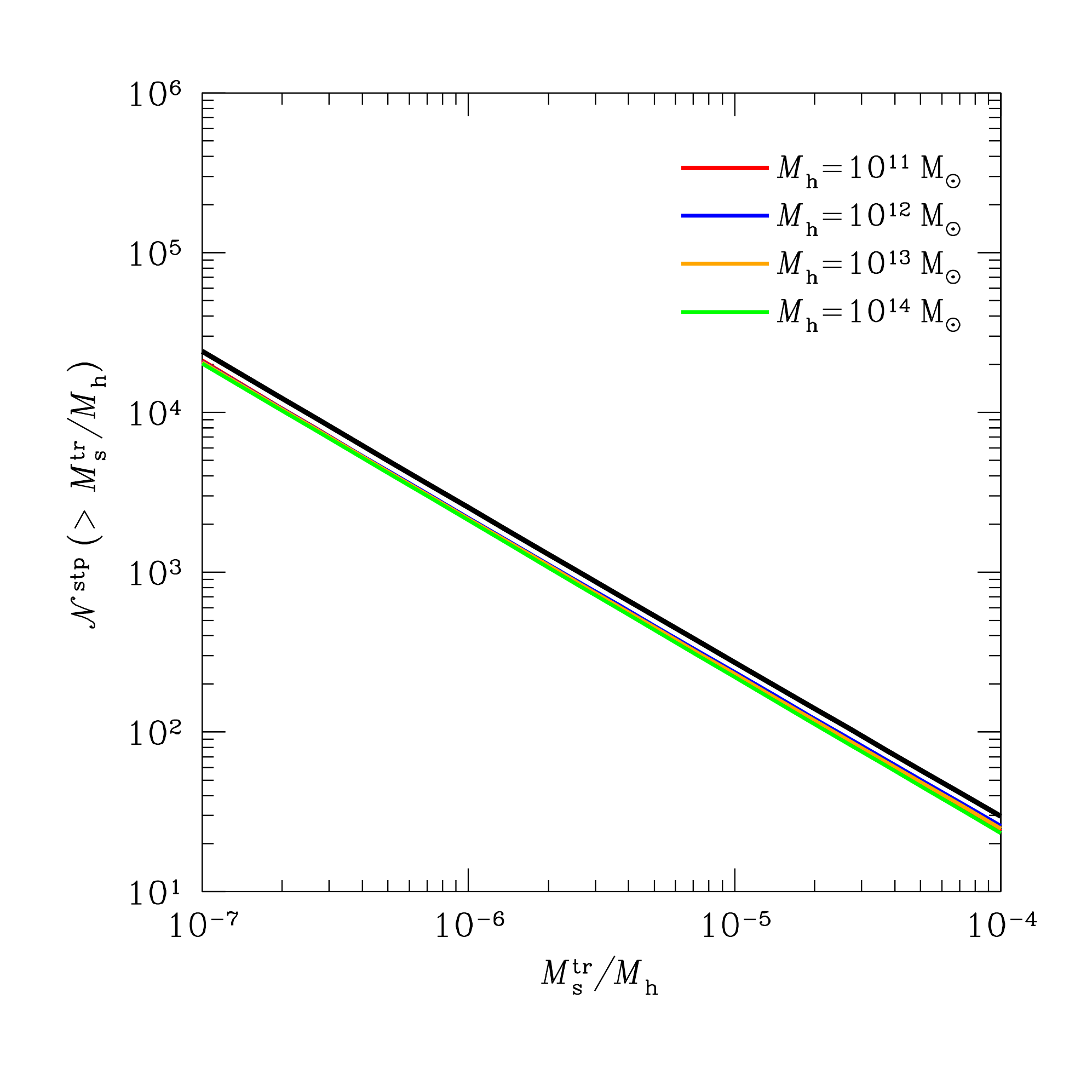}}
\caption{Cumulative subhalo MFs predicted for ordinary haloes of several masses multiplied by $[M\h/(10^{12}$ \modotc$)]^{-0.07}$. The universal MF of purely accreting haloes (thick black line) is just slightly higher.}
(A colour version of this Figure is available in the online journal.)
\label{f7}
\end{figure}

By integrating over $r$ the previous radial abundance of subhaloes, we are led to the differential subhalo MF in ordinary haloes of different masses (eq.~[\ref{MF}]) and by integration over $\clM\tr$ the corresponding cumulative MFs shown in Figure \ref{f7}.  For the reason mentioned when describing the mass integral of $\mu$, the cumulative MF of ordinary haloes averaged over their formation times shows the same dependence on halo mass as purely accreting haloes. (Strictly speaking, they overlap now when they are multiplied by $[M\h/(10^{12}$ \modotc$)]^{-0.07}$. The slight difference in the power index with respect to that of purely accreting haloes ($-0.08$) arises from the distinct formation time PDFs of haloes with different masses.) In addition, the MF of ordinary haloes averaged over their formation times is somewhat smaller than that of purely accreting haloes, as expected from the discussion on the values of $\mu$ near $R\h$. On the contrary, the MF of individual ordinary haloes fully overlap with the MF of purely accreting haloes of the same mass, which explains the result in Figure \ref{f3}. Only the MF of extreme late-forming haloes is substantially lower because of the rapid decrease of their $\mu$ profile when going away from $R\h$. We will comeback to this behaviour of the MF of the latest-forming haloes in next Section. 

\section{Ordinary Haloes of a Fixed mass and Different Formation Times}\label{specific}

The fraction of accreted subhaloes satisfying any desired property in haloes with a fixed mass $M\h$ at $t\h$ formed in an interval $\Delta$ of time around any desired value $t\f$ can be derived using equation (\ref{one2}) with the formation time PDF restricted within that interval, that is 
\beq
f^{t\f,\Delta}_{\rm [M\h,t\h]}(t)=A\,\Pi (t-t\f,\Delta)\, f_{\rm [M\h,t\h]}(t),
\label{convol}
\eeq
where $\Pi (t-t\f,\Delta)$ is the top-hat function of width $\Delta$ around $t\f$ and $A$ is the normalization factor. This way we are led to 
\beqa
F_{\rm [M\h,t\h]}^{t\f,\Delta}(x,m)=\int_0^{t(x)} \der t \, f^{t\f,\Delta}_{\rm [M\h,t\h]}(t)\,F\PA_{\rm [M\h,t\h]}(x,m)~~~~~~~\nonumber\\
+  \int_{t(x)}^{t\h} \der t\, f^{t\f,\Delta}_{\rm [M\h,t\h]}(t) \bar F_{\rm [M(t),t]}[1,m],~~~~~~~
\label{onebis}
\eeqa
Thus, equation (\ref{onebis}) with the function $\bar F_{\rm [M(t),t]}[1,m]$ solution of the integral equation (\ref{one3}) leads to the desired fraction $F^{t\f\pm \Delta/2}_{\rm [M\h,t\h]}(x,m)$. 

This procedure can be applied to infer the $\mu$ and $f\rel$ profiles in ordinary haloes with formation times averaged inside any desired interval. Following \citet{Bea20}, who studied MW-mass haloes assuming the {\it WMAP7} cosmology like we do, we consider two extreme intervals: one with an upper bound at $t/t\h=0.28$ ($z=1.74$), corresponding to $r/R\h=0.36$, that embraces the 20\% earliest-forming objects and another interval with a lower bound at $t/t\h=0.76$ ($z= 0.29$), corresponding to $r/R\h=0.82$, that embraces the 20\% latest-forming objects. (The previous figures correspond to haloes with $M\h=10^{12}$ \modotc; for haloes of different masses we have chosen suited values to delimit the same kind of early- and late-forming objects.) We remark that these two intervals coincide with those used by \citet{Bea20} despite the different halo formation time definition they adopt: the time the halo reaches 50\% of its final mass. Indeed, haloes with $M\h$ at $t\h$ that follow the pure accretion track reaches half the final mass at $z=3.5$. Thus, all haloes undergoing the last major merger after that redshift automatically reach that pure accretion track at the same moment and, hence, they are also seen to form there according to the alternate formation time definition. While haloes undergoing the last major merger before $z=3.5$, in the alternate formation time definition will be seen to form some time after when they will reach $z=3.5$. But, since this redshift is higher than the upper redshift of our interval of the earliest-forming haloes, all these haloes will lie in that interval according to both formation time definitions, even though their individual formation times will differ in both cases.

\begin{figure}
\centerline{\includegraphics[scale=.45,bb= 18 50 560 558]{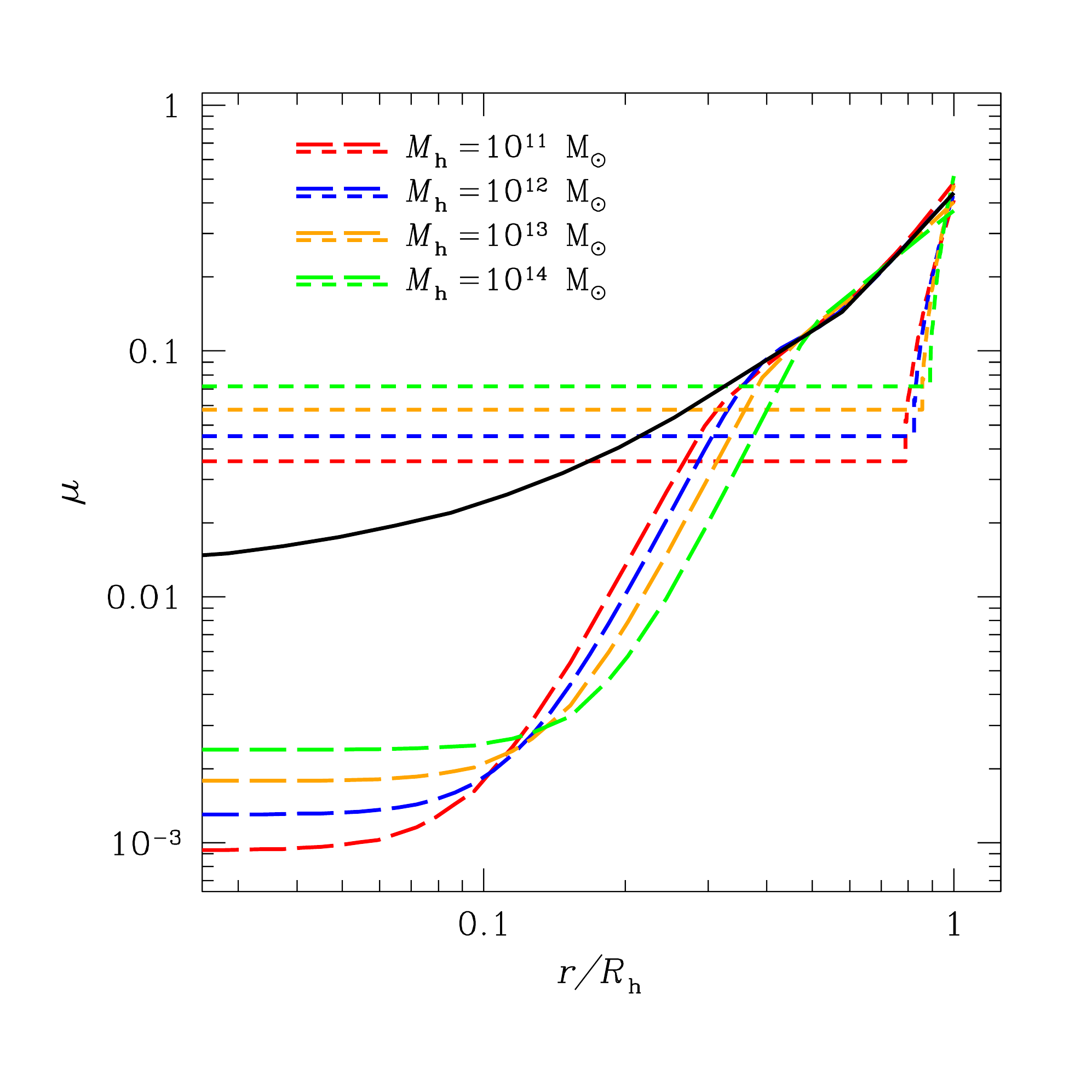}}
\caption{Mean truncated-to-original subhalo mass ratio profiles predicted using the CUSP $M$--$c$ relation for the 20\% earliest-forming haloes (long-dashed lines) and the 20\% latest-forming haloes (short-dashed lines) of several masses.
For comparison we plot the same profile for a purely accreting halo of $10^{12}$ \modot (thick black line).}
(A colour version of this Figure is available in the online journal.)
\label{f8}
\end{figure}

\begin{figure}
\centerline{\includegraphics[scale=.45,bb= 18 50 560 558]{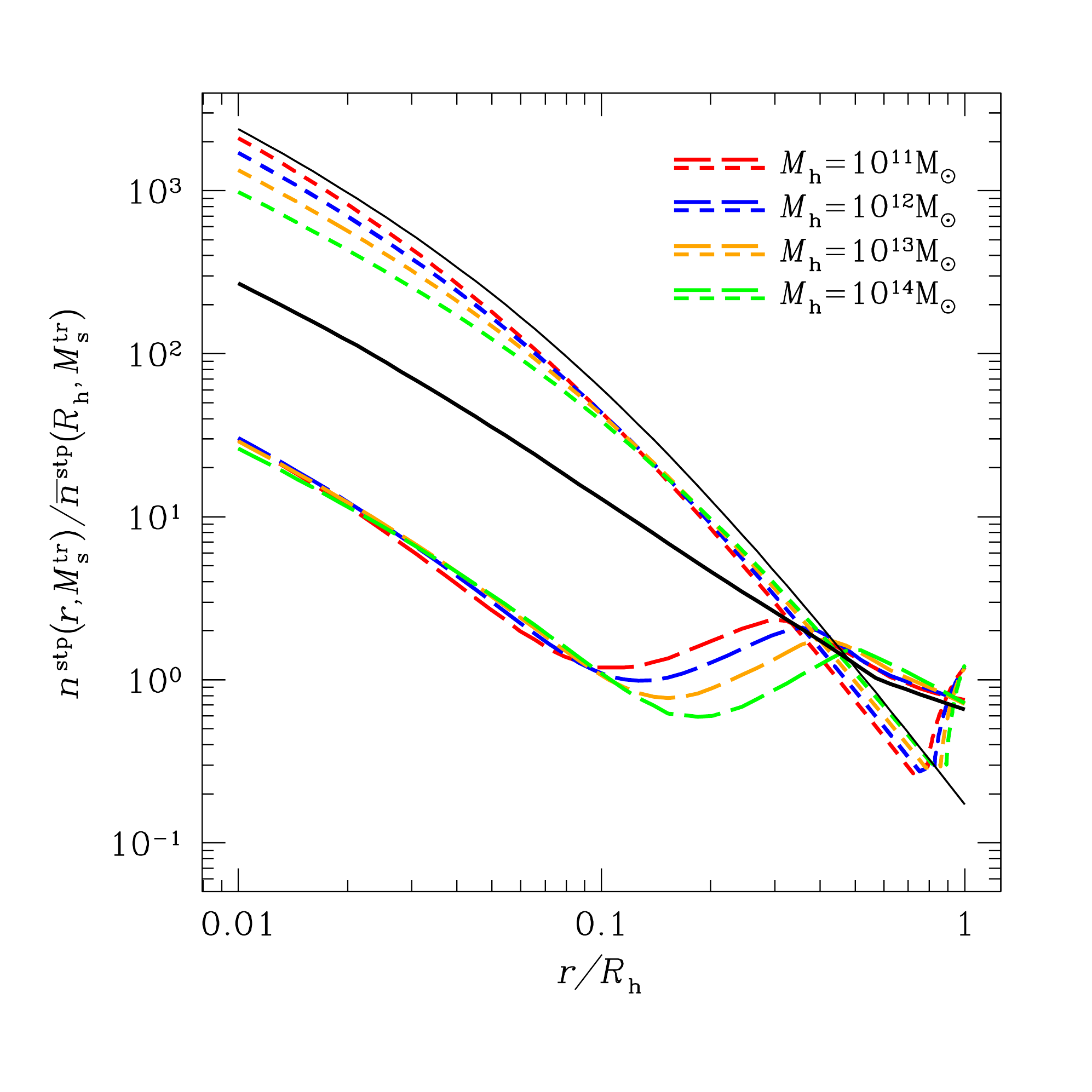}}
\caption{Same as Figure \ref{f8} (same lines) but for the scaled subhalo number density profiles. Also plotted is the formation time-independent mass density profile of haloes with $10^{12}$ \modot (thin black line) scaled so as to facilitate the comparison.}
(A colour version of this Figure is available in the online journal.)
\label{f9}
\end{figure}

The $\mu$ profiles for ordinary haloes formed in those two extreme intervals are shown in Figure \ref{f8}. Their shape in each case is readily understood from our previous explanations of the behaviour of that profile in Section \ref{ordinary}. It also explains the behaviour of the corresponding scaled subhalo number density profiles, $n(r,\clM)/\bar n(\clM)$ (eq.~[\ref{new}]), shown in Figure \ref{f9}, that follow from their radial abundances given by equation (\ref{M}). As can be seen, the mean scaled subhalo number density profiles of the earliest-forming haloes of any mass are quite similar to each other: they are parallel to the scaled mass density profile of their respective haloes until $r/R\h\sim 0.3$, where such haloes typically formed, and then rapidly increase reaching the $\mu$ profile of purely accreting haloes at a finite radius substantially smaller than $R\h$. On the contrary, the scaled profiles of the latest-forming haloes keep their initial trend parallel to the density profile of the respective haloes until a much larger radius, where they suddenly recover to reach the $\mu$ profile of purely accreting haloes at $R\h$. Thus, the two scaled subhalo number density profiles are quite distinct, meaning that this property would be  a good tool for probing the halo formation time. However, since these profiles are scaled to the total number of subhaloes of each mass, they do not inform on the subhalo richness of haloes.

\begin{figure}
\includegraphics[scale=.45,bb= 18 50 560 558]{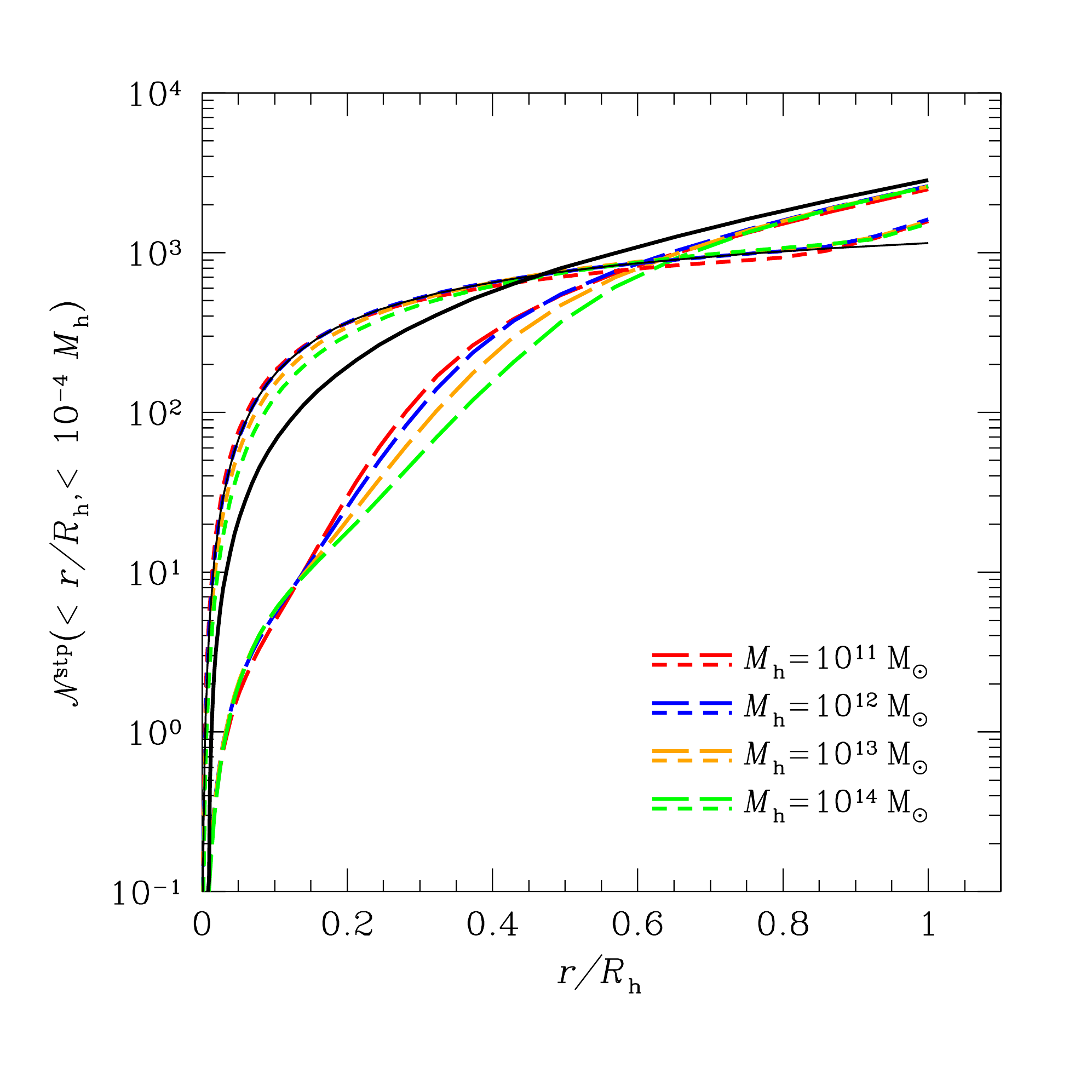}
\caption{Integrated number density profiles of plain subhaloes with masses $10^{-6}M\h < \clM\tr < 10^{-3}M\h$ predicted by CUSP for the 20\% earliest-forming (long-dashed lines) and 20\% latest-forming (short-dashed lines) haloes of different masses $M\h$ multiplied by $[M\h/(10^{12}$ \modotc)]$^{-0.08}$. For comparison we plot the same profile for purely accreting haloes of the MW-mass (thick black line) and the integrated mass density profile of the (purely accreting or ordinary) halo (thin black line) with suited zero-point so as to essentially overlap with the profiles of the latest-forming haloes.}
(A colour version of this Figure is available in the online journal.)
\label{f10}
\end{figure}

\begin{figure}
\includegraphics[scale=.45,bb= 18 50 560 558]{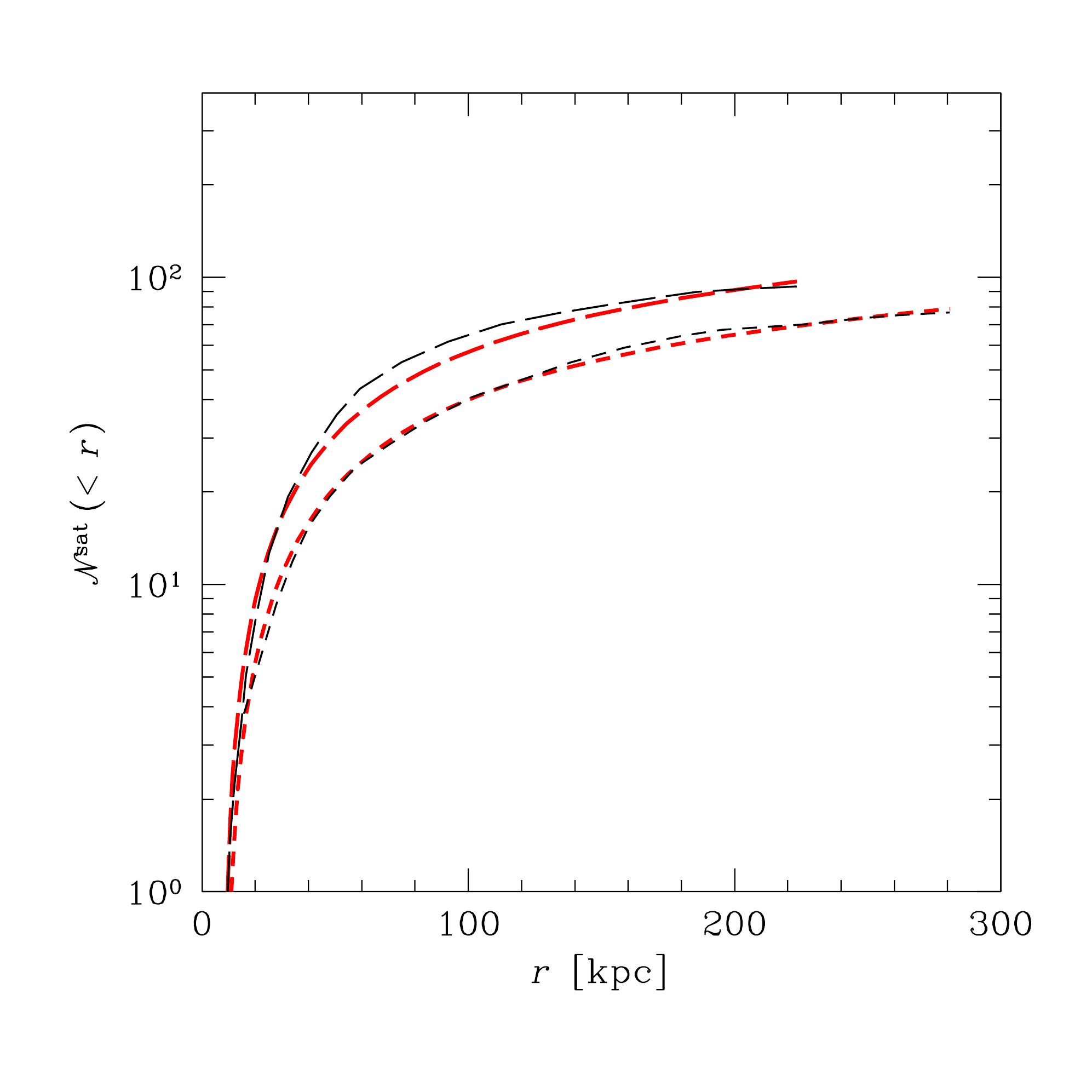}
\caption{Integrated number density profiles of accreted subhaloes with masses $1.4 \times 10^{8}$ \modot\ $<\clM<5\times 10^{8}$ \modot corresponding to satellites with stellar masses $M_\star < 10^{6}$ \modot in haloes with $M_{200}= 1 \times 10^{12}$ \modot (long-dashed red line) and $M_{200}= 1.3 \times 10^{12}$ \modot (short-dashed red line) predicted by CUSP compared to the average profiles of the 20\% earliest-forming (long-dashed black line) and 20\% latest-forming (short-dashed black line) haloes found by \citet{Bea20} for MW-mass haloes ($M_{200}\sim 1-1.3\ \times 10^{12}$ \modotc).}
(A colour version of this Figure is available in the online journal.)
\label{f11}
\end{figure}

An alternative estimate of the radial distribution of subhaloes that is non-scaled and easier to determine in simulations as well as observations because using cumulative quantities is that put forward by \citet{Bea20}, namely the profile of the total number of subhaloes with masses below some given value that lies inside each radius $r$. In Figures \ref{f10} and \ref{f11} we plot two versions of this integrated density profile: one dealing with ``plain" subhaloes as used so far, with masses in the range $10^{-6}M\h < \clM\tr < 10^{-3}M\h$ (the lower limit is supposed to account for the typical mass resolution of simulations of MW-mass haloes), and another one dealing with ``luminous" subhaloes, i.e. subhaloes harbouring faint and ultra-faint satellites (with stellar mass less than $3.2 \times 10^6$ \modotc), ready to be compared with the results of the simulations of MW-mass haloes carried by \citet{Bea20} (see their Fig.~5). 

As can be seen in Figure \ref{f10}, the integrated radial number density profile of plain subaloes in haloes of different masses essentially overlap, in both the 20\% earliest- and 20\% latest-forming haloes, when they are multiplied by $[M\h/(10^{12}$ \modotc)]$^{-0.08}$; there is just a small dispersion at intermediate radii due to the fact that the formation time intervals cannot be scaled to $M\h$. The fact that this mass dependence essentially coincides with that of the subhalo MFs\footnote{It is again $0.08$ rather than $0.07$ due to the narrower formation time interval used here compared to that used in the subhalo MF of ordinary haloes.} is not unsurprising: both properties arise from the integration over $r$ of the original radial abundance of stripped subhaloes (eq.~[\ref{M}]). Regarding the radial distributions in haloes of any fixed mass, we see that in the earliest-forming haloes it is slightly higher at large radii ($r/R\h\ga 0.6$) than in the latest-forming ones. Since subhaloes at those radii contribute the most to the total number of objects in the haloes, this means that the former haloes are richer than the latter ones (see the MFs of the two kinds of haloes below). But this is not the only difference: the profile of the earliest-forming haloes decreases towards the halo centre much more steeply than that of the latest-forming haloes, which essentially stay parallel to the integrated mass density profile of the halo. As a consequence, both profiles cross at $r\sim 0.06R\h$. Thus, the integrated radial distribution of plain subhaloes would be a very good probe of halo assembly history provided subhalo masses could be determined.  

In Figure \ref{f11} we compare the theoretical radial distributions of luminous subhaloes to those of satellites found by \citet{Bea20} in the 20\% earliest- and latest-forming MW-mass haloes ($M_{200}\sim 1 - 1.3 \times 10^{12}$ \modotc). In principle, to calculate such theoretical profiles we should find the upper and lower mass limits of stripped subhaloes that correspond to the upper and lower stellar mass limits of $3.2 \times 10^6$ \modot and zero \modotc, respectively, used by Bose et al. But this correspondence depends on the radius in an unknown way, so that calculation is hard to achieve. Fortunately, there is an alternative way to calculate the desired radial distributions. The vast majority of satellites with stellar masses below $\sim 3.2 \times 10^6$ \modot necessarily formed in haloes with masses between the minimum mass for star formation, $\clM\sim 1.4 \times 10^{8}$ \modot \citep{Bea20}, and the maximum mass of haloes having formed $3.2 \times 10^6$ \modot of stars, $\sim 5\times 10^{8}$ \modot. Certainly, a small fraction of satellites with a final stellar mass slightly below $3.2 \times 10^6$ \modot may have suffered very strong tidal stripping affecting not only dark matter but also stars so that the initial mass of their subhaloes may be slightly larger than $5\times 10^{8}$ \modotc. Consequently, the upper mass limit is actually somewhat fuzzy. Nonetheless, since the radius of the stellar component is typically 1\% of the radius of their host haloes \citep{K13}, the fraction of satellites with initial stellar mass larger than $\mu \clM$ able to lose stars is very small. In addition, any small variation in the upper mass limit has an insignificant effect on the total number of haloes within the bracketed mass range. (Only do variations in the lower mass limit substantially affect that number due to the much higher abundance of objects at the low mass end.) Therefore, the number of accreted subhaloes within that mass range is a very robust estimate for the number of satellites with final masses below $3.2 \times 10^6$ \modotc. And, since low mass subhaloes, unaffected by dynamical friction, stay at the same apocentric radius as when they were accreted, the integrated radial distribution of satellites with masses in the included sample should essentially coincide with that of {\it accreted} subhaloes in the corresponding mass range independent of radius. We have confirmed the robustness of this counting against small changes in the upper subhalo mass limit.

Certainly, the previous reasoning seems to contradict the results of simulations. As discussed in Paper I (see also \citealt{Jea19}), the number density profile of accreted subhaloes of any mass is independent of the formation time of the host halo. Yet, \citet{Bea20} found a significant difference between the radial distributions of satellites in the 20\% earliest- and 20\% latest-forming haloes. The clue for that apparent contradiction is the different formation time PDF of haloes of different masses (Fig.~\ref{f4}). According to it, the 20\% earliest- and 20\% latest-forming haloes should essentially coincide with the less and most massive objects, respectively, in the sample of haloes with masses $M_{200}\sim 1-1.3\ \times 10^{12}$ \modot used by \citet{Bea20}. And, since the abundance of accreted subhaloes within any given mass range is proportional to $M\h^{-1}$ (see Sec.~\ref{PA}), the 20\% earliest-forming haloes should be about $1.3$ times richer than the 20\% latest-forming ones, where $1.3$ is the ratio of extreme halo masses in the sample.

In Figure \ref{f11} the integrated radial distributions of luminous subhaloes or, more exactly, the corresponding accreted subhaloes predicted by CUSP for haloes with the two extreme masses are compared to the integrated radial distributions of satellites (including orphan objects) found by \citet{Bea20} for the 20 \% earliest- and 20 \% latest-forming haloes with $M_{200} \sim 1 - 1.3 \times 10^{12}$ \modotc. These authors provide the results obtained from the {\it Copernicus Complexio Low Resolution} (COLOR) simulation and the {\it Copernicus Complexio} (COCO) suite of simulations, with substantially higher resolution but much poorer statistics (COLOR has $\sim 20$ times more haloes than COCO). In any case, the mass resolution in both simulations is sufficient to detect star formation in all haloes with masses larger than $1.4 \times 10^8$ \modotc, so including orphan satellites in stripped subhaloes with ending masses below the resolution mass should be sufficient to count all satellites. Yet, the higher resolution of the COCO simulations yields twice more satellites of the relevant masses than in the COLOR simulations (see Fig.~1 of \citealt{Bea20}). Since all-level subhaloes are also twice more abundant than first-level ones (Paper I), that result suggests that many satellites in the COCO simulations lie in high-level subhaloes. This would explain why including one orphan satellite per disrupted (first-level) subhalo in the COLOR simulation is not enough to recover the satellite abundance found in the COCO simulations. This conclusion is also supported by the fact that the satellite abundance in the COCO simulations agrees with the (all-level) subhalo abundance found in merger tree Monte Carlo simulations according to the EPS formalism \citep{Bea20}. But the idea that 50\% of all satellites lies in high-level subhaloes is little realistic if we think about MW satellites. We have thus chosen to compare the radial distribution of (first-level) accreted subhaloes predicted by CUSP to the radial distribution of satellites found in the COLOR simulation.

As can be seen in Figure \ref{f11} the predicted profiles almost fully agree, indeed, with those found by \citet{Bea20}. The only slight difference in the profiles of the earliest-forming haloes is likely due to the effects of dynamical friction, excluded from our treatment. The upper subhalo mass limit of $\sim 5\times 10^{8}$ \modot is a factor $\sim 5$ higher than the minimum mass of subhaloes suffering significant dynamical friction, so in early-forming haloes subhaloes could indeed be slightly more concentrated towards the centre. In any event, this agreement gives strong support to the conclusion that the radial distributions of satellites in haloes with masses in a very narrow range around any fixed value do not depend on their formation times. In addition, it reinforces the idea that satellites in the COLOR simulation lie in first-level subhaloes only, while those in the COCO simulations likely also occupy higher-level subhaloes.

\begin{figure}
\centerline{\includegraphics[scale=.45,bb= 18 50 560 558]{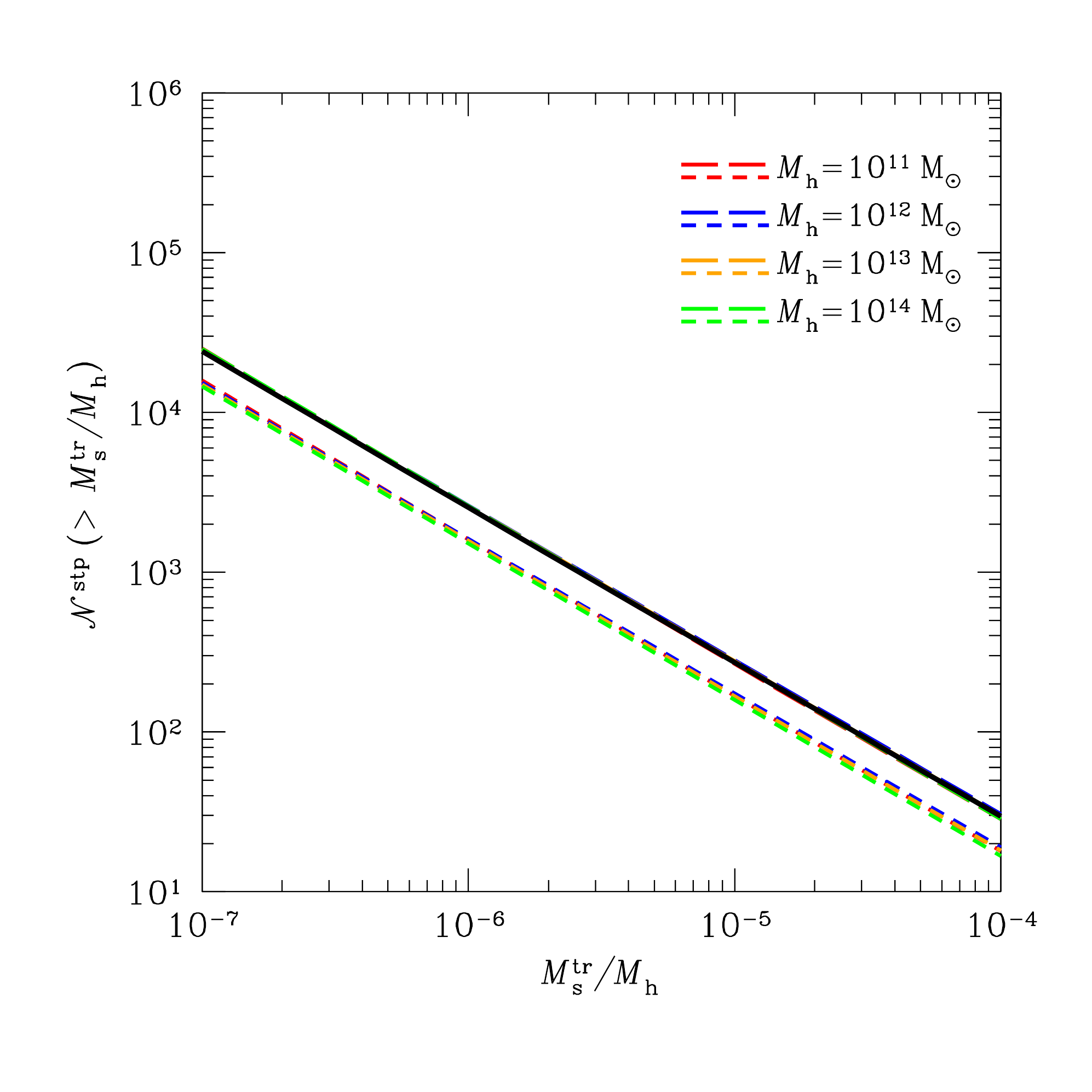}}
\caption{Same as Figure \ref{f10} but for the cumulative subhalo MFs (also multiplied by $[M\h/(10^{12}$ \modotc)]$^{-0.08}$}.
(A colour version of this Figure is available in the online journal.)
\label{f12}
\end{figure}

The inclusion of orphan satellites in simulations is crucial for the previous comparison to be meaningful. Otherwise stripping would affect the radial distribution of satellites through the varying number of disrupted subhaloes produced (e.g. \citealt{Gea19}) and we could not use accreted subhaloes to count satellites. This does not mean, of course, that we can use the radial profiles of {\it non-orphan} satellites to probe halo formation times. Even though this is in principle possible in simulations, it is not in observations: reaching higher magnitudes allows a better determination of the radial distributions (and MFs) of satellites (e.g. \citealt{Gea21}), but it does not alter the fundamental result that the abundance of satellites coincides with that of accreted subhaloes, independent of halo formation time. A better determination of the radial distribution of ultra-faint satellites should be useful, however, in connection with the ``missing satellite problem" itself. By comparing the radial distribution (or the MF) of ultra-faint satellites to the theoretical one(s) of the corresponding low-mass {\it accreted subhaloes} (eq.~\ref{nacc}, with ${\cal N}\acc(\clM)$ given by the well-normalised MF of accreted subhaloes; see Paper I), it should be possible to confirm whether or not there is a problem in the satellite abundance. Note that such a test, unaffected by the complications arising from stripping and dynamical friction, has the added advantage that it is independent of the assembly history of haloes. To do that it would be advisable, however, to exclude the very central region of haloes so as to avoid the effect of satellite destruction by central galaxies (see \citealt{Bea20} and references therein). 

Finally, integrating the radial abundances, ${\cal N}^{t\f\pm \Delta}_{\rm [M\h,t\h]}(r,\clM)$ out to $R\h$, we are led to the differential MF of plain or luminous subhaloes in haloes of different formation times in early-forming and late-forming haloes of any fixed mass. In Figure \ref{f12} we depict the cumulative MFs of plain subhaloes for haloes of different masses. As usual, all MFs depend on halo mass as $M\h^{0.08}$. Regarding their similarity with the respective MF of purely accreting  haloes, we see that, as expected, the MFs in the earliest-forming haloes of any mass overlap with it, while those of the latest-forming haloes are substantially lower. The difference is consistent with that found in the radial distribution of subhaloes at large radii, while the shape of all these MFs is always the same, essentially a power-law of index close to $-1$. In this sense, subhalo MF is a poorer probe of halo formation time than the radial distribution of subhaloes.

\section{SUMMARY AND DISCUSSION}\label{dis}

The present Paper is the last one of a series of three devoted to provide a comprehensive insight on halo substructure. Making use of the CUSP formalism we derived, in Paper I, the properties of accreted subhaloes from the statistics of their seeds (nested peaks) in the primordial Gaussian random density field. In Paper II, we developed a detailed stripping model and used the properties of accreted subhaloes to derive those of stripped ones taking into account the repetitive tidal truncation and shock heating they suffer as they orbit within their hosts. As argued in Paper II, the properties of stripped subhaloes depend on the particular assembly history of their host halo. Nevertheless, their derivation was carried in the simplest, little realistic case of purely accreting haloes. Here we have extended those results to ordinary haloes, i.e. having undergone major mergers, with the aim to: 1) see how the properties of substructure determined in Paper II are modified by the action of major mergers, 2) determine their dependence on substructure on halo mass and formation time (i.e. the time of the last major merger) and 3) find the capabilities of substructure as a probe of halo assembly history.

In a formal plane, it has been shown that all the properties of substructure in haloes of any mass and formation time are encoded in the mean truncated-to-original subhalo mass ratio profile, $\mu(r,\clM\tr)$, weakly dependent, actually, on $\clM\tr$. This profile is flat inside the radius reached by the object at its last major merger (inside which the system was completely scrambled) and rapidly approaches, at larger radii, the $\mu$ profile of the purely accreting halo of the same mass, setting the correspondence between $r$ and $t$ which can be used to determine the time of the merger. This profile is at the base of all the remaining more practical results.

Regarding our first goal, we have demonstrated that the agreement found in Paper II between our predictions for purely accreting haloes and the results of simulations dealing with ordinary ones was not casual. Even though major mergers affect the properties of substructure in ordinary haloes with respect to purely accreting ones, their subhalo MF remains essentially unaltered. Major mergers only leave a substantial imprint in the radial distribution of subhaloes.

These results have important repercussions on our second goal. On the one hand, the dependence on halo mass of the subhalo MF is the same in ordinary haloes as in purely accreting ones. Since in the latter the formation time plays no role, that dependence can only arise from the mass dependence on mass of halo concentration, as found in Paper II. This dependence on halo mass is already seen, of course, in the (non-scaled) radial distribution of subhaloes whose integral over the radius leads to the MF. But the radial distribution of subhaloes in haloes of any given mass has the added interest that it also harbours information on the formation time of the host halo. Indeed, the scaled subhalo number density profiles in ordinary haloes, independent of subhalo mass, is steeper than in purely accreting haloes, being proportional to the density profile of the halo inside the scrambling radius reached at its formation. This scaled version of the radial distribution of subhaloes does not inform on the subhalo richness, but another non-scaled version of it, the integrated subhalo number density profile, does. We have shown that this latter profile is higher at large radii contributing the most to the total subhalo abundance in early-forming haloes than in late-forming ones. Moreover, the profiles in the two kinds of haloes are also very different as they cross each other at some intermediate radii. Of course, the different richness of haloes of a given mass is also reflected in their subhalo MF, which in extremely late-forming haloes is found to be substantially lower than in all the rest. 

All these results lead to the following conclusion regarding the third goal: the radial distribution of subhaloes is very useful to probe the halo formation time, much better than the subhalo MF. Unfortunately, this conclusion only holds for plain subhaloes, not for satellites. Indeed, the properties of satellites do not depend on how their host halo stripped subhaloes. They only depend on the masses of accreted subhaloes where they formed, whose properties are independent of the halo formation time. Thus, the properties of ultra-faint satellites do not inform on the formation time of the host halo. The good news is that, by simply comparing their radial distribution to that of very low mass subhaloes they should unambiguously tell whether or not there really is a ``missing satellite problem", without depending on the particular formation history of the host halo. 

In its current form, our analytic treatment does not account for dynamical friction. Of course, this is not a drawback when dealing with low enough mass subhaloes (faint enough satellites). But, if we want to deal with more massive subhaloes (more luminous satellites), we should account for that process. That possibility would be very welcome because, even though the radial distribution of satellites does not depend on subhalo stripping, it is sensitive to dynamical friction. Thus, the properties of satellites could still inform on the halo assembly history through the effect of dynamical friction. Work in this line is currently in progress \citep{Sea21}.

\vspace{0.75cm} \par\noindent
{\bf ACKNOWLEDGEMENTS} 

\vspace{11pt}\noindent
One of us, I.B., has benefited of a MEXT 
scholarship by the Japanese MECSST. This work was funded by grants CEX2019-000918-M (Unidad de Excelencia `Mar\'ia de Maeztu') and PID2019-109361GB-100 (together with FEDER funds) by MCIN/AEI/10.13039/501100011033 and by the grant 2017SGR643 funded by the Catalan DEC.

\par\vspace{0.75cm}\noindent
{\bf DATA AVAILABILITY}

\vspace{11pt}\noindent 
The data underlying this article will be shared on reasonable request to the corresponding author.

{}

\end{document}